\documentclass[preprint,onecolumn,nofootinbib]{revtex4}
\pdfoutput=1
\usepackage{float}
\usepackage{amsmath}
\usepackage{amsfonts}
\usepackage{amssymb,ulem}
\usepackage{color}
\usepackage{dcolumn}
\usepackage{caption2}
\usepackage{subfigure}

\usepackage{MnSymbol,wasysym}
\usepackage{braket,diagbox}
\usepackage{eurosym}
\usepackage{calrsfs}
\usepackage[usenames,dvipsnames,svgnames]{xcolor}

\newcommand{\RNum}[1]{\uppercase\expandafter{\romannumeral #1\relax}}
\usepackage[colorlinks=true,linkcolor=blue,urlcolor=blue,filecolor=black,citecolor=red,
pdfstartview=FitV,pdftitle={},pdfsubject={},pdfkeywords={},pdfpagemode=None,bookmarksopen=true]{hyperref}
\usepackage{graphicx}

\begin{document}
\baselineskip=0.5 cm

\title{Shadow of Kerr black hole surrounded by a cloud of strings in Rastall gravity and constraints from \text{M87*}}
\author{Qi Sun}
\author{Yu Zhang}
\email{zhangyu@kust.edu.cn}\thanks{(Corresponding author)}
\author{Chen-Hao Xie}
\author{Qi-Quan Li}
\affiliation{Faculty of Science, Kunming University of Science and Technology, Kunming, Yunnan 650500, China}

\begin{abstract}
\baselineskip=0.45 cm
Motivated by the first image of a black hole captured by the Event Horizon Telescope (EHT), there has been a surge of research using observations of black hole shadows to test theories of gravity. In this paper, we carry out a study related to the shadow of Kerr black holes surrounded by a cloud of strings in Rastall gravity, which deviates from the Kerr black hole due to the presence of the string parameter $a_0$ and the parameter $\beta$. The horizons, ergospheres, and photon region of the black hole are shown. Moreover, we explore the shadow and observations of the black hole, which are closely linked to the parameters $a_0$ and $\beta$. By treating \text{M87*} as a Kerr black hole surrounded by a cloud of strings under Rastall gravity, we constrain the black hole parameters using the EHT observations. For a given $\beta$, the circularity deviation of the black hole obeys $\Delta C\lesssim0.1$ in all regions.  The angular diameter $\theta_{d}=42\pm3\mu as$ provides the upper bound of parameters $a$ and $a_0$ for fixed $\beta$. The shadow axis ratio satisfies the observation data of EHT ($1<D_x\lesssim4/3$) in the whole space for a given $\beta$. These results are consistent with the public information from EHT. In other words, candidates for real astrophysical black holes can be Kerr black holes surrounded by a cloud of strings in Rastall gravity.

\end{abstract}


\maketitle

\tableofcontents
\section{Introduction}
Since Synge and Luminet \cite{Synge:1966okc} proposed an expression for the angular radius of the photon capture region around the Schwarzschild black hole, research on black hole and wormhole shadows has been booming. Afterwards, Bardeen \cite{Bardeen} discovered that the spin would distort the shape of the Kerr black hole shadow, which differs from the perfect circle of the Schwarzschild black hole shadow. Wei and Liu \cite{Wei:2020ght} investigated the shadow of the rotating Gauss-Bonnet black hole and used the astronomical data to restrict the black hole parameters. Liu et al. \cite{Liu:2023oab} explored the shadow of a Kerr-like black hole surrounded by a dark matter halo and compared it with the Kerr black hole and then obtained the upper limit of shape parameter of Einasto density profile. Furthermore, with the development of the study on the shadow of rotating black holes, it is increasingly understood that trajectories of photons around black holes are strongly associated with the geometry of black holes \cite{Falcke:1999pj,Shen:2005cw,Amarilla:2010zq,Yumoto:2012kz,Atamurotov:2013sca,Xu:2018mkl,Bambi:2019tjh,Allahyari:2019jqz,Li:2020drn,Atamurotov:2021hck,Hou:2021okc,Sun:2022wya,Singh:2017xle,Zubair:2023cep,Zubair:2023cor}. Therefore, physicists can obtain details about the near-horizon properties of black holes by observing their shadows. With the rise of the study of black hole shadows, the shadows of wormholes have also received much attention \cite{Tinchev:2013nba,Abdujabbarov:2015pqp,Ohgami:2016iqm,Abdujabbarov:2016efm,Shaikh:2018kfv,Gyulchev:2018fmd,Vagnozzi:2019apd,Stuchlik:2019uvf,Long:2019nox,Konoplya:2019fpy,Xavier:2020egv,Chen:2020qyp,Pantig:2020uhp,Khodadi:2020jij,Jusufi:2020mmy,Javed:2020mjb,Chowdhuri:2020ipb,Khan:2020ngg,Lee:2021sws,Shaikh:2021yux,Gussmann:2021mjj,Junior:2021dyw,Konoplya:2021slg,Chaudhary:2021uuk,Pantig:2021zqe,Wang:2021art,Kumar:2017tdw,Adler:2022qtb,Neto:2022pmu,Pulice:2023dqw,Zahid:2023csk,Wang:2023vcv,Meng:2023wgi,Meng:2023htc,Sengo:2022jif,Karmakar:2023mhs,Singh:2023zmy,Jha:2023nkh,Hoshimov:2023tlz,Wu:2024hxr}. The authors of Ref. \cite{Nedkova:2013msa} discussed the influence of spin on the shadow of rotating wormholes and compared the wormhole shadow with the Kerr black hole shadow. They found the difference between their shadows increases at large spin parameters. It is worth noting that a recent work \cite{Wang:2020emr} discusses the shadow of the asymmetric thin-shell wormhole model whose impact parameters of null geodesics are usually discontinuous. The study revealed that the shadow sizes of this model rely on the photon sphere in the other side of the spacetime. In the study of black hole shadows, Hioki and Maeda \cite{Hioki:2009na} presented the radius of the reference circle ($R_S$) and distortion ($\delta _S$) to determine the spin parameters and inclination angle based on the shape of Kerr black hole shadow for the sake of extracting and analyzing information from black hole shadow. Then Tsukamoto et al. \cite{Tsukamoto:2014tja} stated a method to distinguish the shadows of Kerr black hole and other rotating black holes. However, the flaw of the two methods is that they require shadow boundary having some symmetry to calculate the observables accurately. Later, to characterize a haphazard shadow shape, Kumar and Ghosh \cite{Kumar:2018ple} defined two observables: shadow area $(A)$ and oblateness $(D)$ to describe the black hole shadow. By calculating these parameters and comparing them with EHT observations, we can estimate the black hole parameters in general relativity (GR) and modified theories of gravity (MoG) such as spin, mass and charge.

Black holes are expected to leave a shadow on the plane of the observer when transparent emission region surrounds it. The shadow results from the gravitational light bending and the photons are captured at the event horizon. In 2019, the EHT collaboration \cite{EventHorizonTelescope:2019dse,EventHorizonTelescope:2019uob,EventHorizonTelescope:2019jan,EventHorizonTelescope:2019ths,EventHorizonTelescope:2019pgp,EventHorizonTelescope:2019ggy} photographed the first shadow image of the supermassive black hole \text{M87*} captured by a global very long baseline interferometry array. This image makes the black hole shadow become physical reality beyond theoretical predictions \cite{Meng:2022kjs}. Unveiling the image simultaneously, the EHT also released information about it: deviation of circularity $\Delta C\lesssim0.1$; oblateness $1<D_x\lesssim4/3$; asymmetric bright emission ring with angular diameter about $\theta_{d}=42\pm3\mu as$. These observations are consistent with the shadow of Kerr black holes predicted in GR, but they are not sufficient to rule out Kerr or non-Kerr black holes in MoG. In this way, we can use the observations of M87* to explore the viability of black holes in MoG, while constraining black hole parameters and even distinguish black holes under different gravities.

As one of the MoG, Rastall gravity is proposed with the assumption that the covariant divergence of the energy-momentum tensor is non-zero \cite{Rastall:1972swe}. The model of Rastall gravity might account for the problem of dark energy and dark matter. It appears to be consistent with observational data \cite{Al-Rawaf:1995xkt} such as the age of the universe, Hubble parameters, and helium nucleosynthesis. Along with the development of Rastall gravity theory, there is also controversy, that is, whether Rastall gravity is equivalent to Einstein theory of gravity. Visser \cite{Visser:2017gpz} believed that Rastall gravity is just a rearrangement of the matter part of Einstein gravity theory, rendering the two theories entirely equivalent. Darabi et al. \cite{Darabi:2017coc} disagreed with the view of Visser and thought that Rastall gravity is not equal to Einstein gravity. They argued that the definition of the energy-momentum tensor of Rastall gravity is incorrect in the paper of Visser. Moreover, they pointed out that Rastall gravity is a more open theory and more consistent with observations than GR. Regardless of whether Rastall gravity is equal to Einstein gravity, it is still being tested by astronomical observations. So research in Rastall gravity continues to flourish. It is not limited to the solution of compact objects in Rastall gravity, the quasinormal modes of black holes, and black hole shadows \cite{Lin:2018coh,Lin:2018dgx,Ovgun:2019jdo,Shao:2020gwr,Abbas:2020kju,Shahzad:2020gjj,Lobo:2020jfl,Vagnozzi:2022moj,Ali:2022ziu,Li:2022wzi,Shao:2022oqv,Afrin:2022ztr}. Recently, Malik et al. \cite{Malik:2024iii} investigated the stability analysis of anisotropic stellar structures in Rastall theory of gravity. Atazadeh and Hadi \cite{Atazadeh:2023wdw} studied the source of the black bounce in Rastall gravity.

In string theory, it is believed that the point-like particles of particle physics are replaced by one-dimensional objects called strings \cite{Herscovich:2010vr}. Letelier \cite{Letelier:1979ej} pioneered the idea that the source of the gravitational field may be a cloud of strings. Simultaneously, he gave an exact solution for the Schwarzschild black hole surrounded by a collection of strings in GR. Since then, a cloud of strings viewed as a source of gravitational field has gradually come into sight of researchers. Toledo and Bezerra \cite{Toledo:2019szg} obtained the solution of the black hole with a cloud of strings in pure Lovelock gravity. In 2020, Cai and Miao \cite{Cai:2019nlo} derived the solution of the Schwarzschild black hole surrounded by a cloud of strings in Rastall gravity and calculated its quasinormal modes and spectroscopy. On this basis, Li and Zhou \cite{Li:2020zxi} applied Newman-Janis algorithm to get the solution of Kerr black hole surrounded by a cloud of strings in Rastall gravity. Rodrigues and Silva \cite{Rodrigues:2022rfj} explored the effect of string cloud surrounding Bardeen or Simpson-Visser black holes on their singularity. They also found Bardeen solution becomes singular while the Simpson-Visser solution remains regular when a cloud of strings exists. Yang et al. \cite{Yang:2023agi} derived a rotating black hole mimicker surrounded by the string cloud and studied its shadow. They discovered that the string cloud parameter strongly influences the shadow boundary of the black hole. Besides, they also analyzed the quasinormal modes of black bounces surrounded by a string cloud.

The Kerr black hole surrounded by a cloud of strings in Rastall gravity is an interesting attempt to use a cloud of strings as gravitational field source in MoG. In fact, MoG were proposed to address some flaws in GR. By studying the shadow of black holes in MoG and correlating them with some astronomical data from the EHT, this may open a new window on the study of MoG.

This paper is organized as follows. In Section~\ref{rn}, we consider the Kerr black hole surrounded by a cloud of strings in Rastall gravity and explore the effects of black hole parameters on the horizon structure. Section~\ref{photon} gives the null geodesic equations and obtains the photon region. In Section~\ref{bh shadow}, the black hole shadow at infinity is shown. We study some observables of the black hole shadow and calculate the energy emission rate in Section~\ref{observables}. Section~\ref{m87} is devoted to treating \text{M87*} as a Kerr black hole surrounded by a cloud of strings in Rastall gravity, and constraining the black hole parameters from EHT observations. The last section is our summary of the full paper. For simplicity, we set $G=c=M=1$, unless otherwise specified.

\section{Kerr black hole
surrounded by a cloud of strings in Rastall gravity}\label{rn}

The field equations of Rastall gravity can be written as
\begin{equation}\begin{aligned}
&G_{\mu\nu}+\beta g_{\mu\nu}R=\kappa T_{\mu\nu},\\
&T^{\mu\nu}{}_{;\mu}=\lambda R^{,\nu},
\end{aligned}\end{equation}
where the Rastall dimensionless parameter $\beta=\kappa\lambda$. Here, $\kappa$ denotes the Rastall gravitational coupling constant and $\lambda$ represents the Rastall parameter.

The two-dimensional world sheet $\Pi$ of a string can be written as $x^\alpha = x^\alpha(X^a)$. The action of string is as follows:
\begin{equation}
I_S=\mu\int_{\Pi}\sqrt{-\chi}dX^0dX^1=\mu\int_{\Pi}\sqrt{-\frac{1}{2}\Pi_{\alpha\beta}\Pi^{\alpha\beta}}dX^0dX^1,
\label{action}
\end{equation}
in which $\mu$ denotes a dimensionless constant related to the tension of string. $\chi$ represents the determinant of induced metric
\begin{equation}
\chi_{ab}=g_{\alpha\beta}\frac{\partial x^\alpha}{\partial X^a}\frac{\partial x^\beta}{\partial X^b}.
\end{equation}

$\Pi_{ab}$ has following form:
\begin{equation}
\Pi^{\alpha\beta}=\epsilon^{ab}\frac{\partial x^\alpha}{\partial X^a}\frac{\partial x^\beta}{\partial X^b}.
\end{equation}

\begin{figure*}
    \centering
        \includegraphics[width=0.435\textwidth]{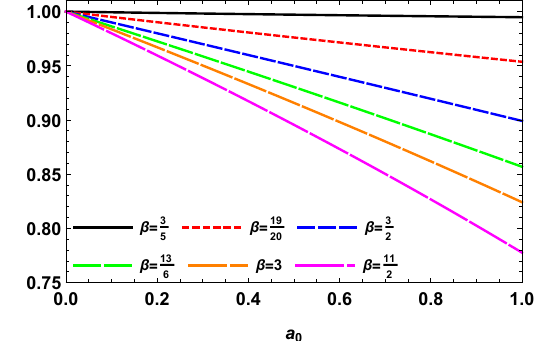}\quad
        \includegraphics[width=0.43\textwidth]{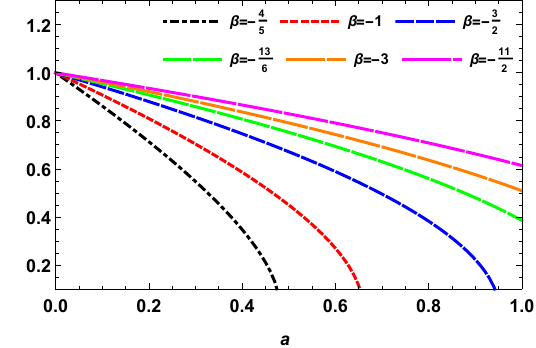}
    \caption{The parameter space $(a_{0}-a)$ of the Kerr black hole surrounded by a cloud of strings in Rastall gravity when $M=1$. The curves denote the extremal black holes distinguishing black holes from naked singularities.}
            \label{dha}
\end{figure*}
And the Levi-Civita symbol $\epsilon^{ab}$ obeys $\epsilon^{01} = -\epsilon^{10} = 1$. Based on Eq.~(\ref{action}), the energy-momentum tensor of the string can be expressed as
\begin{equation}
T^{\alpha\beta}=\mu\frac{\Pi^{\alpha\gamma}\Pi_{\gamma}^{\beta}}{\sqrt{-\chi}}.
\end{equation}

For a collection of strings, the energy-momentum tensor is $T_{\mathrm{cloud}}^{\alpha\beta}=\rho_{s}T^{\alpha\beta}$ ($\rho_{s}$ denotes the number density). Combined with the above and considering the spherically symmetric distribution of string cloud, the metric of Schwarzschild black hole is
\begin{equation}
\mathrm ds^2=-f(r)\mathrm dt^2+f^{-1}(r)\mathrm dr^2+r^2\mathrm d\theta^2+r^2\sin^2\theta\mathrm d\phi^2,
\end{equation}
in which
 \begin{equation}
f(r)=1-\frac{2M}{r}+\frac{4a_{0}(\beta-\frac{1}{2})^2}{(8\beta^2+2\beta-1)r^{\frac{4\beta}{2\beta-1}}},
\end{equation}
where $M$ denotes the black hole mass. $a_0$ represents the string parameter which is related to the energy density of a cloud of strings.

Building on the findings of Ref.~\cite{Cai:2019nlo}, Li and Zhou \cite{Li:2020zxi} obtained the solution of Kerr black holes surrounded by a cloud of strings in Rastall gravity from the Schwarzschild black hole surrounded by a cloud of strings in Rastall gravity via Newman-Janis algorithm. The method is a general procedure to get a stationary and axially symmetric spacetime from static and spherically symmetric spacetime. As a rotating black hole in an extended theory of gravity, the metric describes the Kerr black hole when the gravitational effects of strings serve as basic objects in Rastall gravity. In Boyer-Lindquist coordinates, its line element can be written as

\begin{equation}\begin{aligned}
&
ds^{2}=-\left(1-\frac A\Sigma\right)dt^2+\frac\Sigma\Delta dr^2+\Sigma d\theta^2-\frac{2aA\sin^2\theta}\Sigma dtd\phi\\
&+\sin^2\theta\left(r^2+a^2+\frac{a^2A\sin^2\theta}\Sigma\right)d\phi^2,
\end{aligned}\end{equation}
in which
\begin{equation}\begin{aligned} &A=2Mr-4a_0\lambda_1r^{2-\lambda_2}, \\
&
\Delta=r^2-2Mr+a^2+4a_0\lambda_1r^{2{-}\lambda_2}, \\
&\lambda_1=\frac{\left(\beta-\frac{1}{2}\right)^2}{8\beta^2+2\beta-1},\\
&
\lambda_2=\frac{4\beta}{2\beta-1},\\
&
\Sigma=r^{2}+a^{2}\cos^{2}\theta,\\
\end{aligned}\end{equation}where $a$ represents rotation parameter. When $\beta=0$, the Kerr black hole surrounded by a cloud of strings in Einstein gravity is restored. It will revert to the Kerr black hole in GR when $a_{0}$ vanishes. And the metric will reduce to the Schwarzschild black hole case when both $a$ and $a_{0}$ are zero. Focusing on the asymptotically flat case in our work, we set $\lambda_{2} \ge 0$, i.e., $\beta \le 0$ or $\beta \ge \frac{1}{2}$. Additionally, the metric requires $\beta \neq \pm \frac{1}{2}$.

When $\Sigma \neq 0$ and $\Delta =0$, the line element is singular at spacetime points corresponding to the horizons. And the horizons are determined by
\begin{equation}
\begin{aligned}
\Delta_r=r^2-2Mr+a^2+4a_0\lambda_1r^{2-\lambda_2}=0.
\end{aligned}\end{equation}

Employing a numerical method to get the solution of $\Delta_r =0$, we discover that there are three situations of the roots based on the choice of $a$, $a_0$ and $\beta$. That is, there exist two distinct roots, a double root, or no real positive value. They also correspond to different types of black hole solutions: black holes with event horizon ($r_+$) and Cauchy horizon ($r_{-}$), an extremal black hole with $r_{+}=r_{-}$, and no black hole scenarios.  In Fig. \ref{dha}, we draw the curves of the $(a_0-a)$ plane with the degenerate condition of horizons ($\Delta_r=0$) for different values of $\beta$. These curves represent the case of extremal black holes which distinguish the black hole region and the naked singularity region.

\begin{figure*}
\centering
\includegraphics[width=0.322\textwidth]{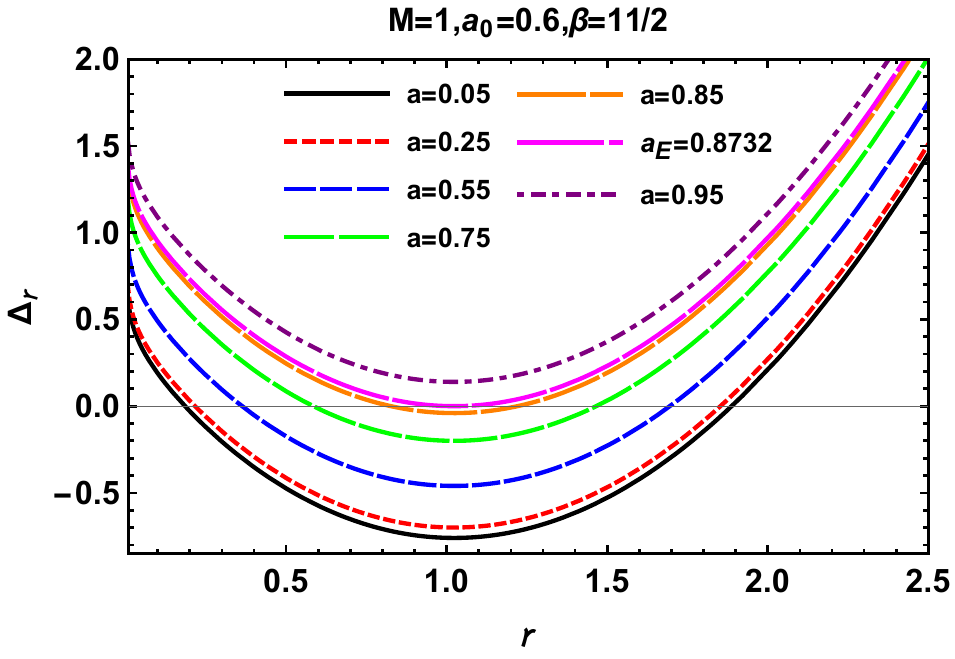}
\includegraphics[width=0.31\textwidth]{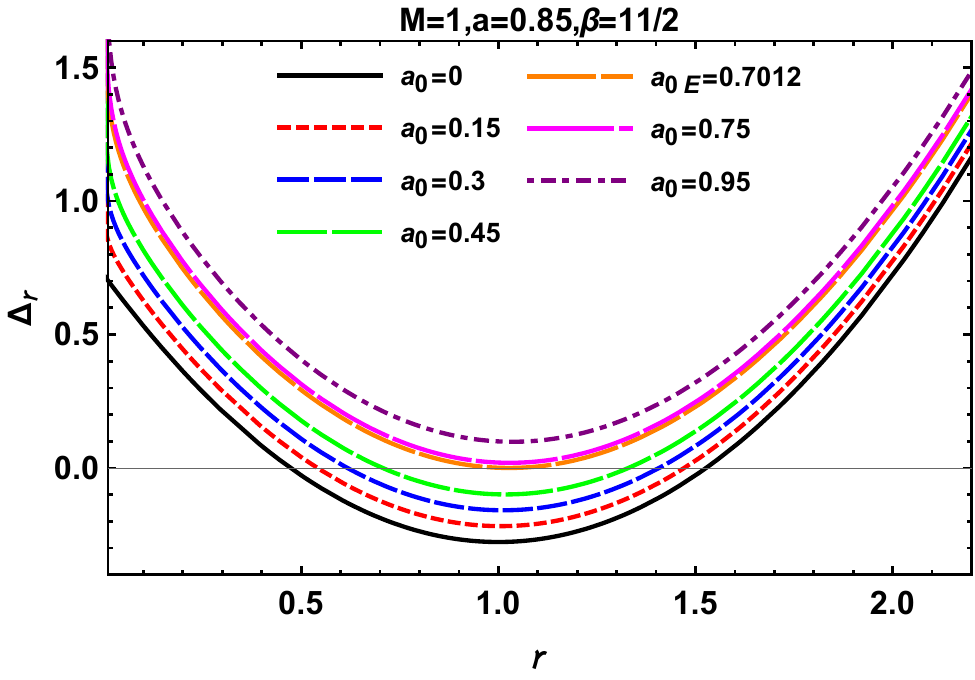}
\includegraphics[width=0.322\textwidth]{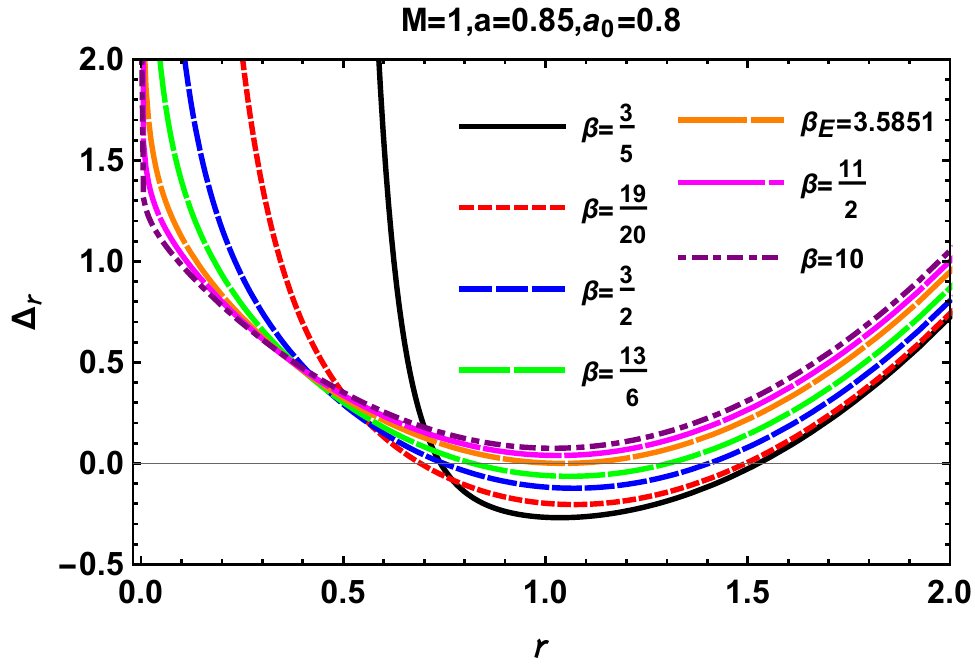}
  \hfill
\includegraphics[width=0.322\textwidth]{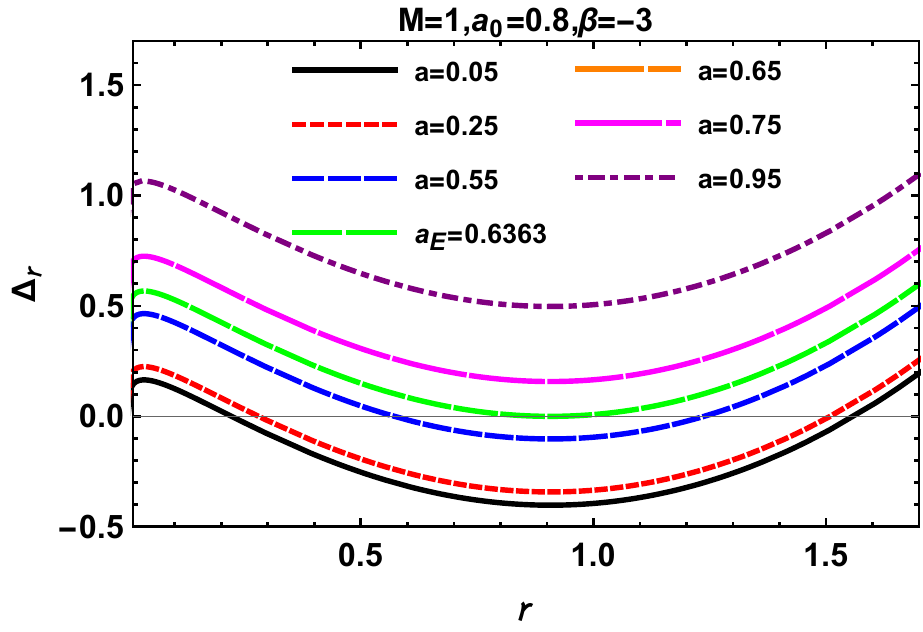}
\includegraphics[width=0.322\textwidth]{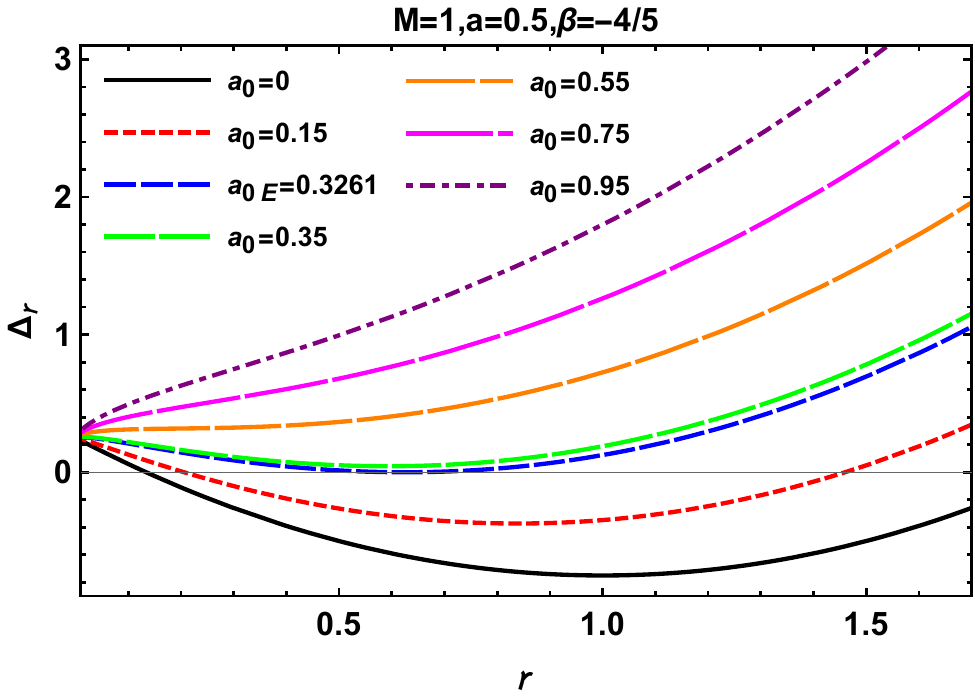}
\includegraphics[width=0.322\textwidth]{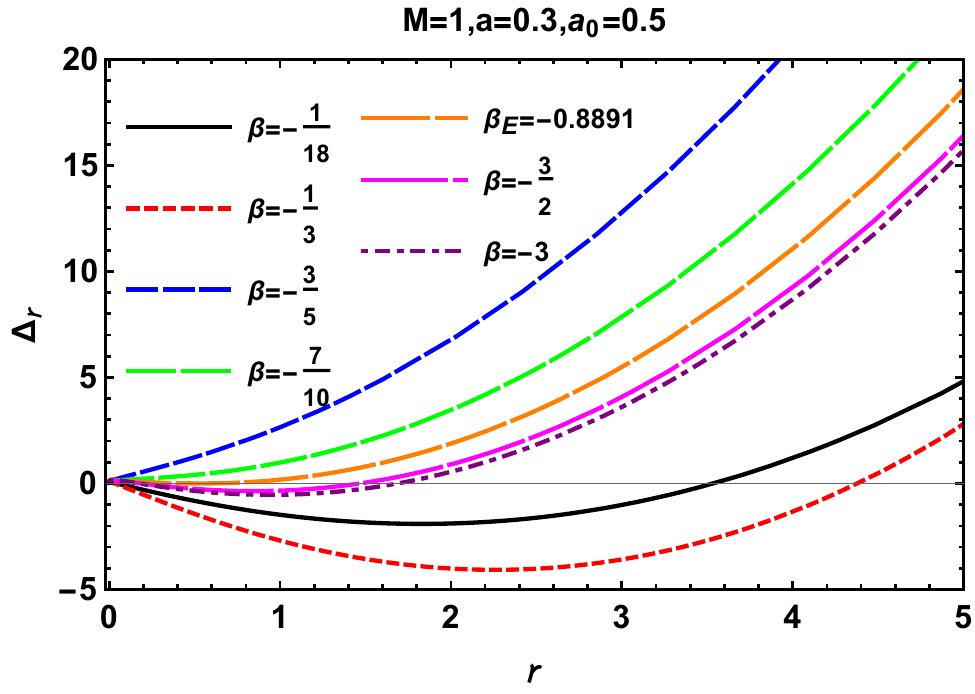}
    \caption{Radial profiles of $\Delta_r$ are shown with varying $a$, $a_{0}$ and $\beta$ of Kerr black hole surrounded by a cloud of strings in Rastall gravity. And the upper row plots the situation of $\beta>0$, while the bottom row draws the situation of $\beta<0$.}
            \label{derta}
\end{figure*}

For the horizon structure of the black hole, we show it in Fig. \ref{derta}. The upper row depicts the case of $\beta>0$, while the second row depicts the case of $\beta<0$. It is natural to know that there is critical extremal value of $a_{E}$ for fixed $a_{0}$ and $\beta$. Given $a$ and $\beta$, critical extremal value of $a_{0E}$ exists. Similarly, the critical extremal value of $\beta_{E}$ exists for setting $a$ and $a_{0}$. From Fig. \ref{derta}, it can be seen that there is a black hole with an event horizon and a Cauchy horizon when $a <a_{E}$, and $ r_{-}$ increases but $ r_{+}$ decreases with the increase of $a$. Whereas $a >a_{E}$ indicates the existence of naked singularity. And the critical extremal value is $a_{E}=0.8732$, $a_{E}=0.6363$ for $a_{0}=0.6$, $\beta=\frac{11}{2} $; $a_{0}=0.8$, $\beta=-3$, respectively.
Likewise, we can see that the Kerr black hole surrounded by a cloud of strings in Rastall gravity exists when $a_{0} <a_{0E}$ and the Cauchy horizon radius increases but the event horizon radius decreases as the value of $a_{0}$ increases. The metric implies a naked singularity when $a_{0}>a_{0E}$. And $a_0=0$ implies Kerr black holes, so Kerr black holes have the smaller Cauchy horizon radii and the bigger event horizon radii. Besides, the critical extremal value $a_{0E}$ depends on $a$ and $\beta$, such as $a_{0E}=0.7012$; $a_{0E}=0.3261$ for $a=0.85$, $\beta=\frac{11}{2}$; $a=0.5$, $\beta=-\frac{4}{5}$, respectively.

Particularly noteworthy is the radial profile of $\Delta_r$ for different $\beta$. We identify critical extremal values: $\beta_{E}=3.5851$ for $a=0.85$ and $a_{0}=0.8$, and $\beta_{E}=-0.8891$ for $a=0.3$ and $a_{0}=0.5$. We first discuss the situation of $\beta>0$. When $\beta<\beta_{E}$, this black hole exists and the event horizon radius keeps decreasing, while the Cauchy horizon radius first decreases and then increases as positive $\beta$ increases. There is no black hole in the region of $\beta>\beta_{E}$. When $\beta<0$, the behavior of $\Delta_r$ with different $\beta$ is interesting. Specifically, in the region from the $\beta_{E}$ to the discontinuity point $\beta=-\frac{1}{2}$, the naked singularity exists. In the rest of the regions, black holes exist.

\section{Null geodesics and photon regions} \label{photon}

Photons from the luminous background or the accretion disk surrounding the black hole are usually unstable when they orbit outside the event horizon. A slight perturbation can cause the photons to be trapped inside the event horizon or escape to infinity after reaching the minimum distance from the black hole. For an observer at infinity, the latter case would confine the photon ring which depends on the parameters associated with the black hole spacetime geometry and is irrelevant to the astrophysical details of the accretion flow model. Therefore, the study of shape and size of the photon ring can be a favorable way to explore the parameters of black holes, which carry the information of black holes. Now, starting from the geodesics of the photon, we aim to obtain the shadow of the Kerr black hole in Rastall gravity with the gravitational effects of strings as basic objects.

The Lagrangian in tensorial form is as follows:
\begin{equation}
\mathcal{L}=\frac{1}{2} g_{\mu\nu} \dot{x}^\mu
\dot{x}^\nu,
\end{equation}
and the generalized momenta can be written as
\begin{equation}
p_\mu=g_{\mu\nu}\dot{x}^\nu.
\end{equation}

Through the above two equations and the metric coefficient, the energy $E$ and the $z$ component angular momentum $L_{z}$ are given by

\begin{equation}\begin{aligned}
&E := -\frac{\partial\mathcal{L}}{\partial\dot{t}}=-g_{tt}\dot{t}-g_{\phi t}\dot{\phi}=p_{t},
\\&
L_{z} := \frac{\partial\mathcal{L}}{\partial\dot{\phi}}=g_{\phi t}\dot{t}+g_{\phi\phi}\dot{\phi}=p_{\phi},
\end{aligned}\end{equation}
in which
\begin{equation}
\dot{x}=\frac{\partial x}{\partial\lambda},
\end{equation}
where $\lambda$ is the affine parameter. We introduce the Hamilton-Jacobi equation, which determines the geodesics of a given spacetime geometry,
\begin{equation}
-\frac{\partial S}{\partial \lambda}=\frac{1}{2} g^{\mu \nu} \frac{\partial S}{\partial x^\mu} \frac{\partial S}{\partial x^\nu}.
\label{h}
\end{equation}

The Jacobi action is
\begin{equation}
S=\frac{1}{2} m_{p}^{2} \lambda-E t+L_{z} \varphi+S_{r}(r)+S_{\theta}(\theta),
\label{s}
\end{equation}
where $m_{p}$ is the mass of test particles.

Through a series of operations, we obtain the null geodesic equations of the test particle ($m_{p}=0$) around the Kerr black hole surrounded by a cloud of strings in Rastall gravity, which can be written as
\begin{eqnarray}
 \Sigma \frac{d t}{d \lambda}&=& \frac{r^2+a^2}{\Delta}\left[E\left(r^2+a^2\right)-a L_z\right]\nonumber\\
 &-&a\left(a E \sin ^2 \theta-L_z\right), \label{t}\\
 \Sigma \frac{d r}{d \lambda}&=&\sqrt{R}, \label{r}\\
 \Sigma \frac{d \theta}{d \lambda}&=&\sqrt{\Theta}, \label{theta}\\
 \Sigma \frac{d \phi}{d \lambda}&=&\frac{a}{\Delta}\left[E\left(r^2+a^2\right)-a L_z\right]-\left(a E-\frac{L_z}{\sin ^2 \theta}\right),\label{fai}
\end{eqnarray}
in which
\begin{equation}
\begin{aligned}
& R(r)=\left[E\left(r^2+a^2\right)-a L_z\right]^2-\Delta\left[m_{p}^2r^2+K\right], \\
&\Theta(\theta)=(K-a^2m_{p}^2\cos^2\theta)-(L_z\csc\theta-aE\sin\theta)^2,
\end{aligned}
\end{equation}

where $K$ is the modified Carter constant. Our main concern is the photon region. For the sake of analyzing the trajectories of photons conveniently, two dimensionless parameters are given,
\begin{equation}
L_E \equiv \frac{L_z}{E}, \quad K_E \equiv \frac{K}{E^2}.
\end{equation}

Obviously, the bounded null geodesics is made up of spherical orbits that are constrained to spheres with $r=constant$. The spherical orbits satisfy $\dot{r}=0$ and $\ddot{r}=0$, which can be determined by
\begin{equation}
R(r)=0, \quad R^{\prime}(r)=0
\label{rc}
\end{equation}
and
\begin{equation}
\Theta(\theta) \geq 0 \quad \text { as } \quad \theta \in[0, \pi].
\end{equation}

By solving Eq.~(\ref{r}) and Eq.~(\ref{rc}), we find that the constants of motion can be written as
\begin{equation}
K_E=\frac{16 r^2 \Delta_r}{\left(\Delta_r^{\prime}\right)^2}, \quad L_E=a+\frac{r^2}{a}-\frac{4 r \Delta_r}{ a\Delta_r^{\prime}}.
\label{kele}
\end{equation}

 \begin{figure*}
    \begin{minipage}{\textwidth}
       \setlength{\leftskip}{1cm}
        \includegraphics[width=0.45\textwidth]{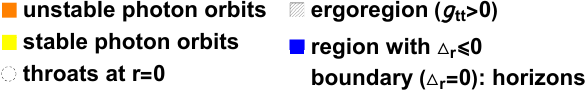}
    \end{minipage}
    \vspace{0.3em}
    \begin{minipage}{\textwidth}
        \centering
\includegraphics[width=0.225\textwidth]{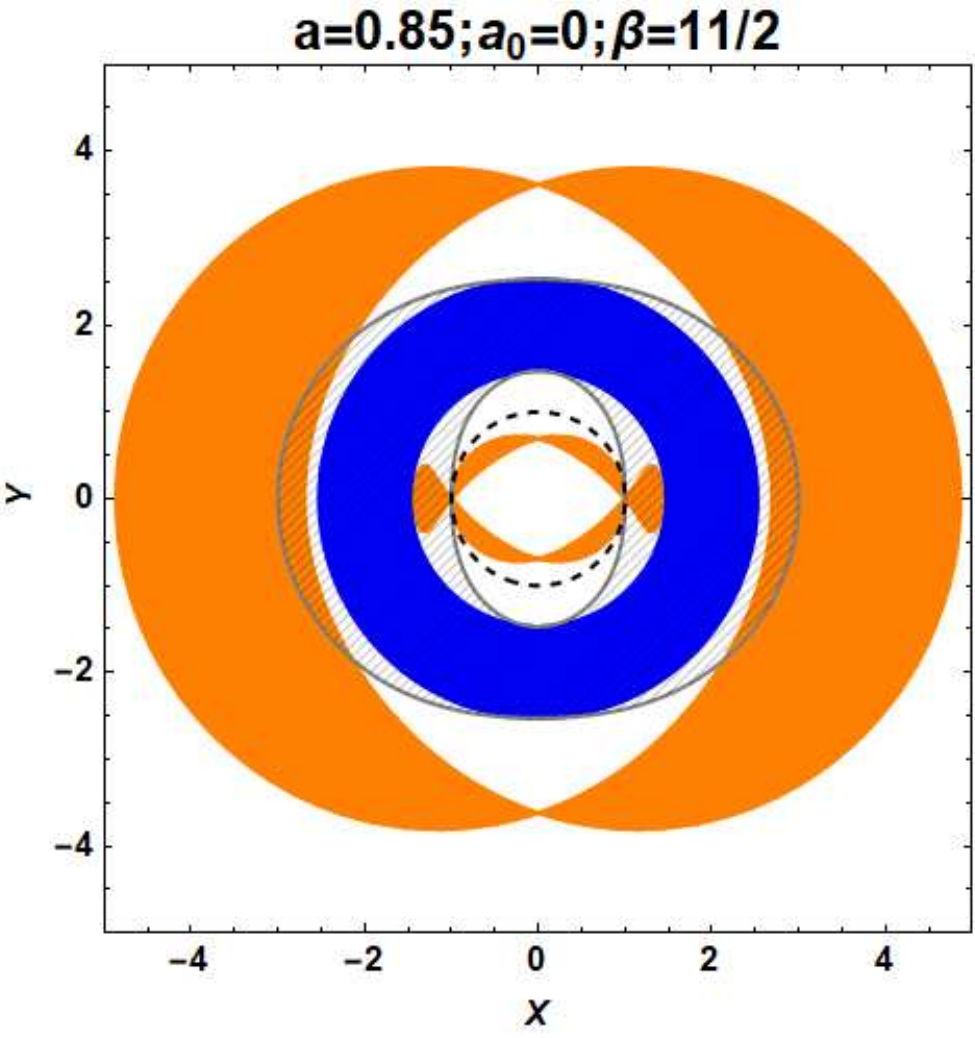}
\includegraphics[width=0.225\textwidth]{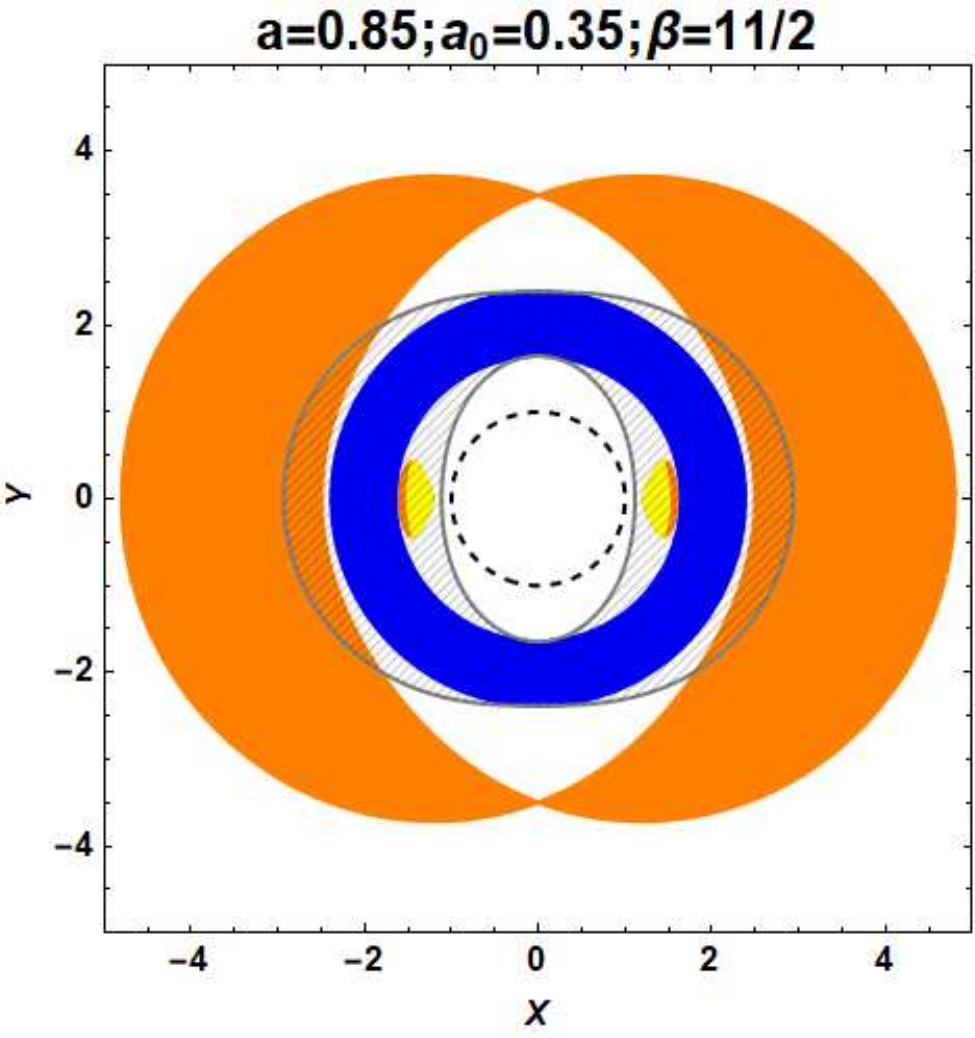}
\includegraphics[width=0.225\textwidth]{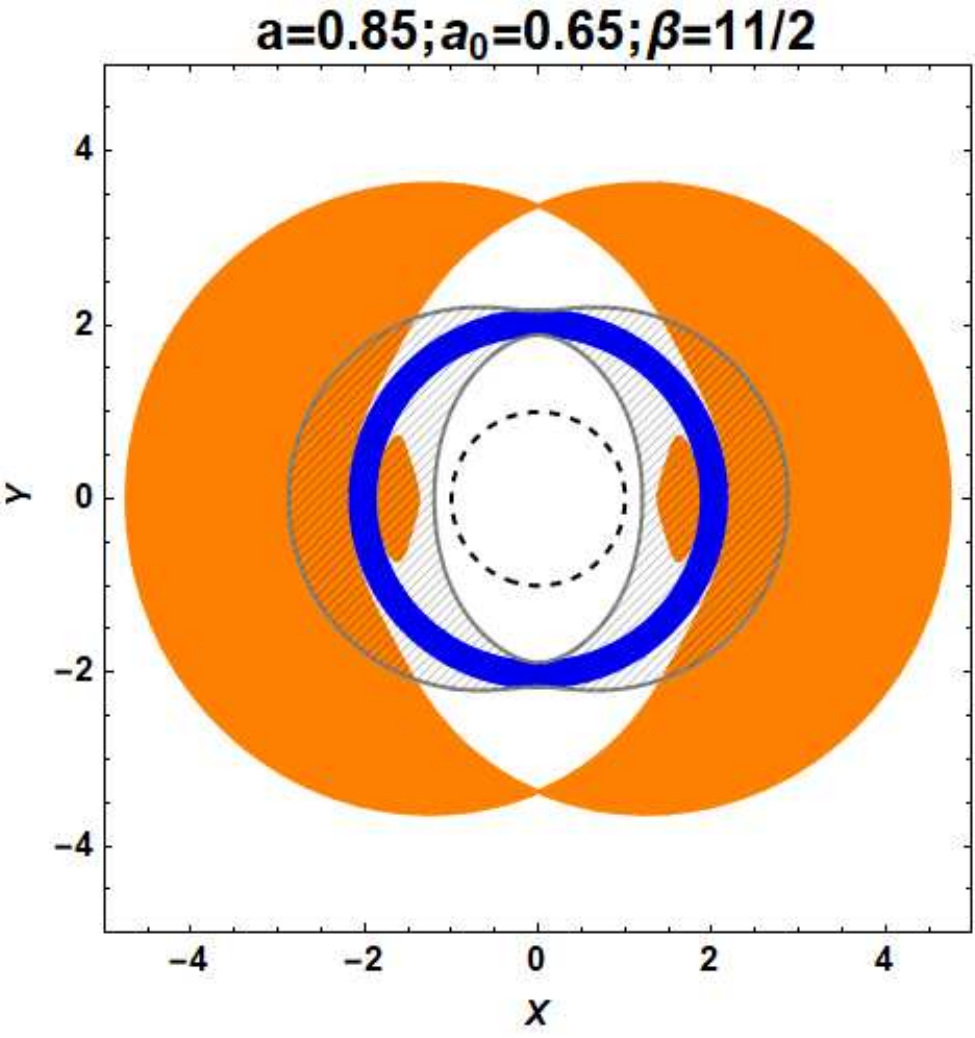}
 \includegraphics[width=0.225\textwidth]{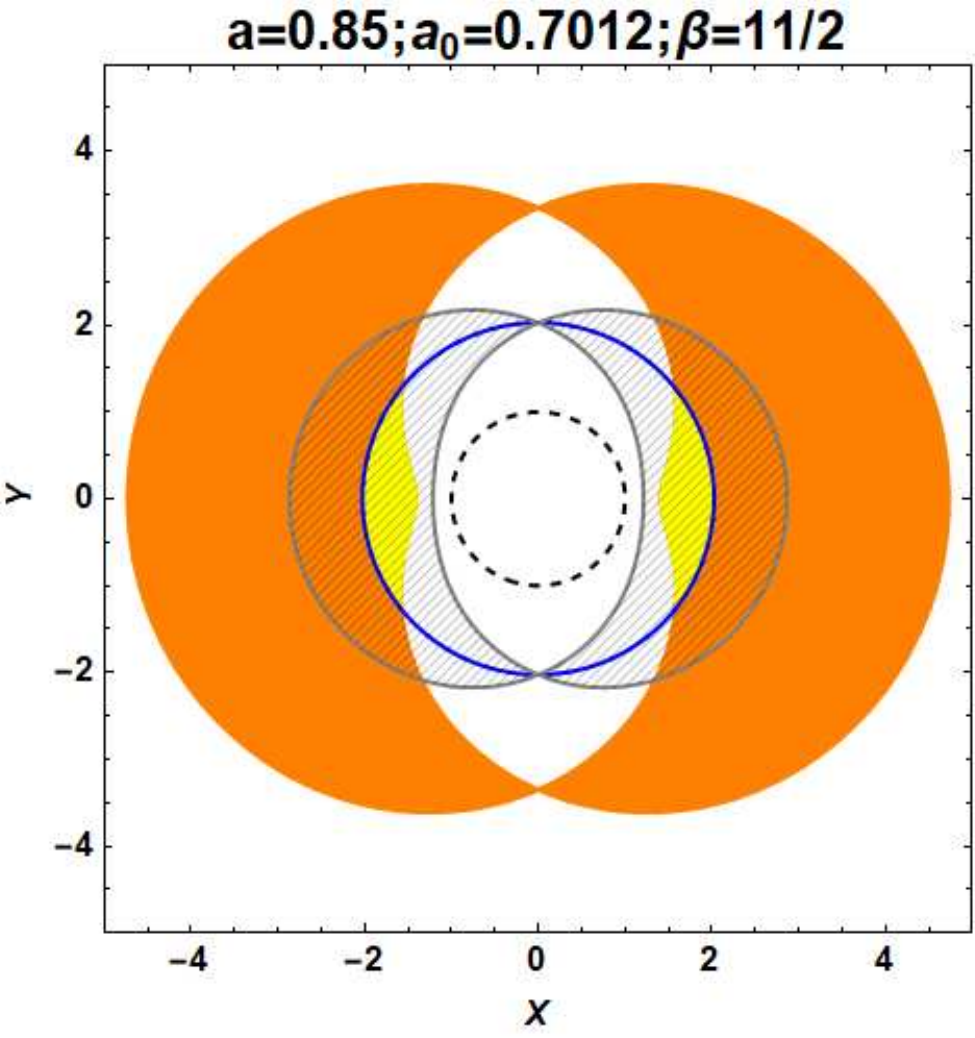}
  \hfill
\includegraphics[width=0.225\textwidth]{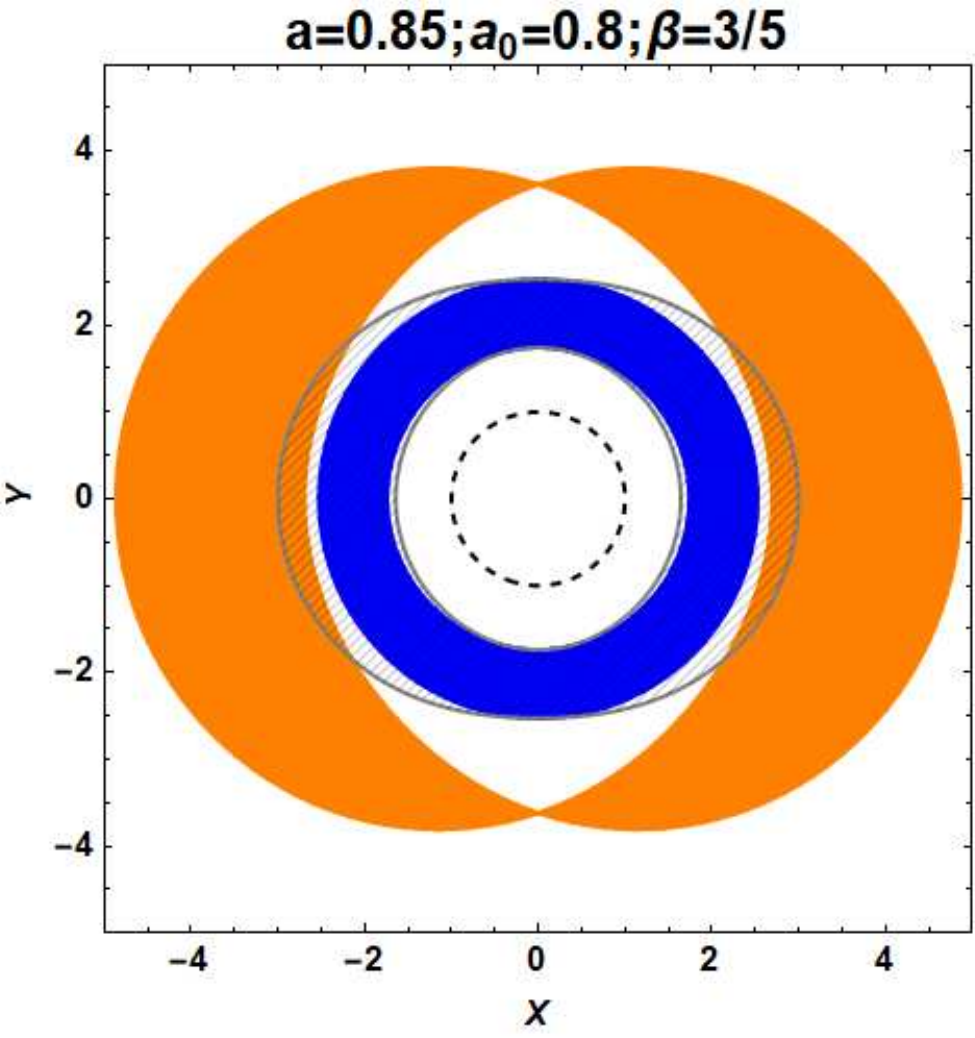}
\includegraphics[width=0.225\textwidth]{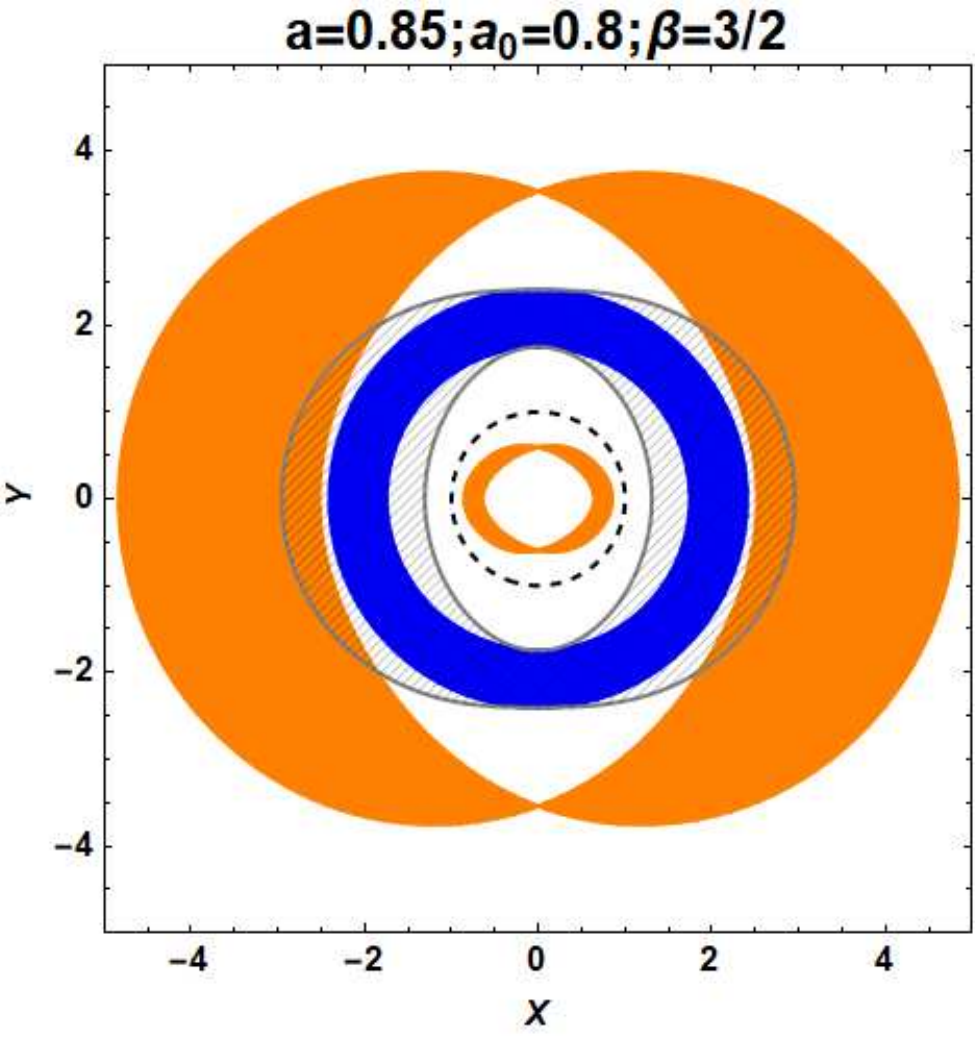}
 \includegraphics[width=0.225\textwidth]{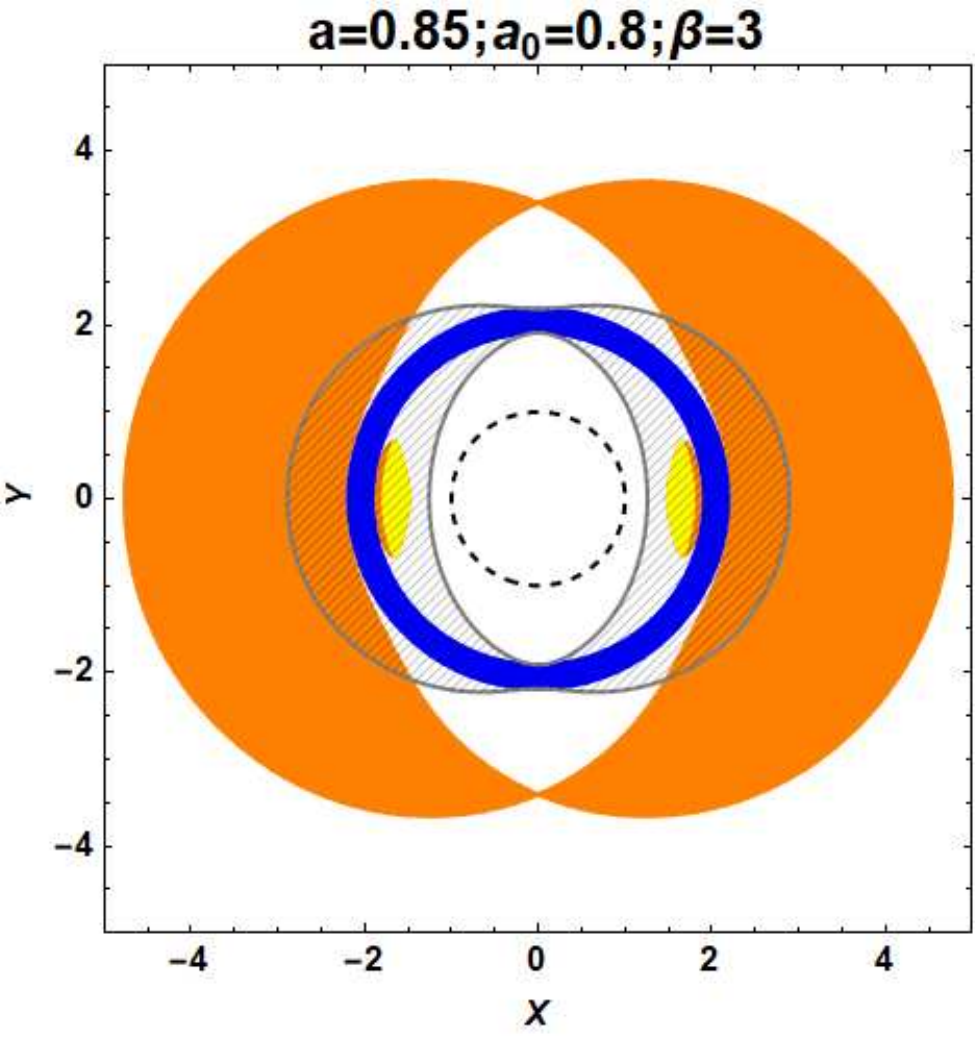}
  \includegraphics[width=0.225\textwidth]{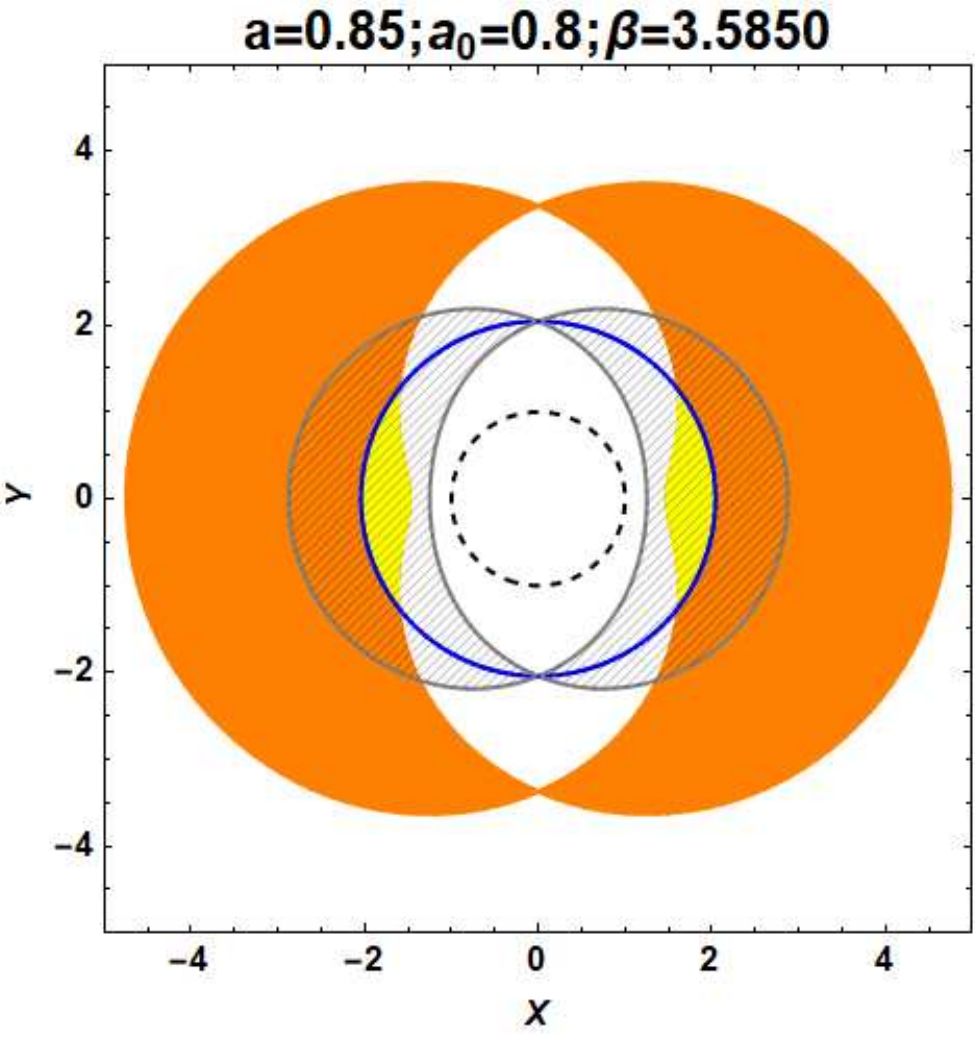}
   \caption{When $\beta>0$, the photon regions and ergosphere region of Kerr black hole surrounded by a cloud of strings in Rastall gravity.}
       \label{dc}
    \end{minipage}
\end{figure*}
\begin{figure*}
\centering
\includegraphics[width=0.225\textwidth]{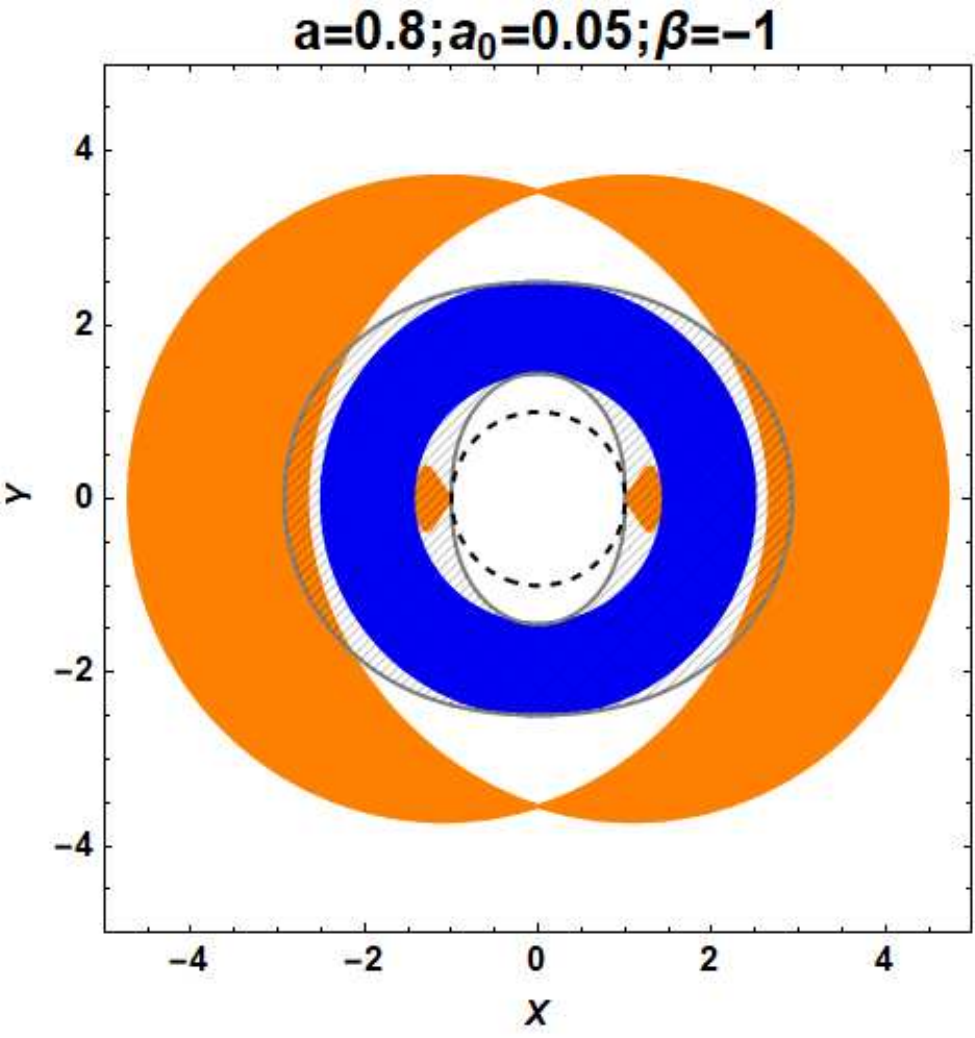}
\includegraphics[width=0.225\textwidth]{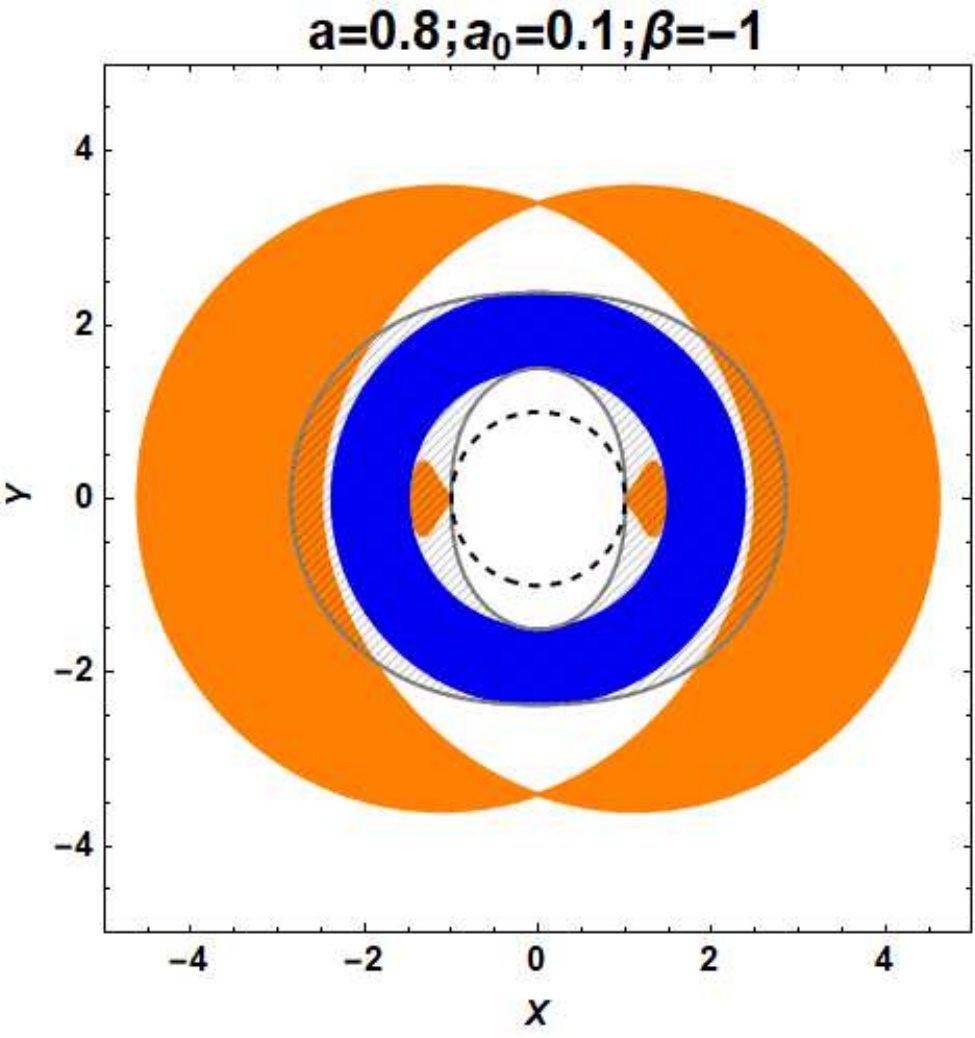}
\includegraphics[width=0.225\textwidth]{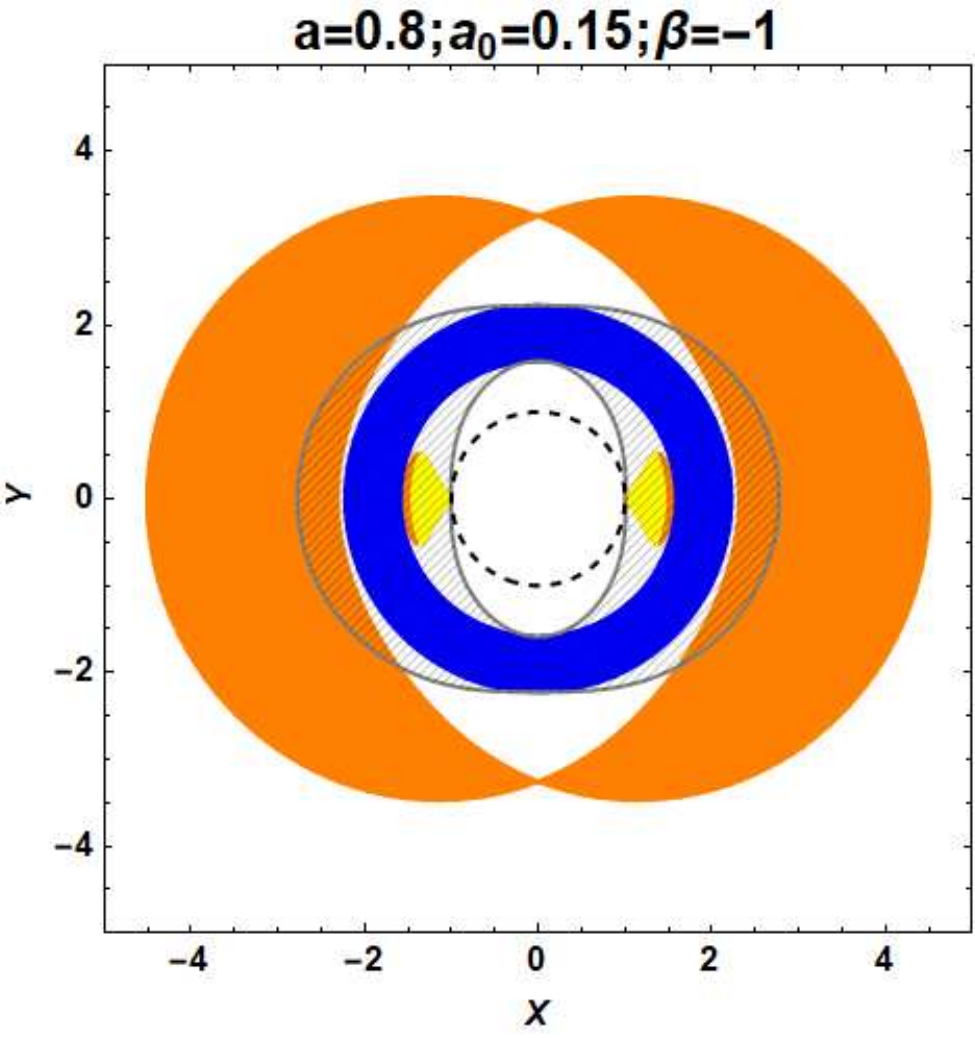}
 \includegraphics[width=0.225\textwidth]{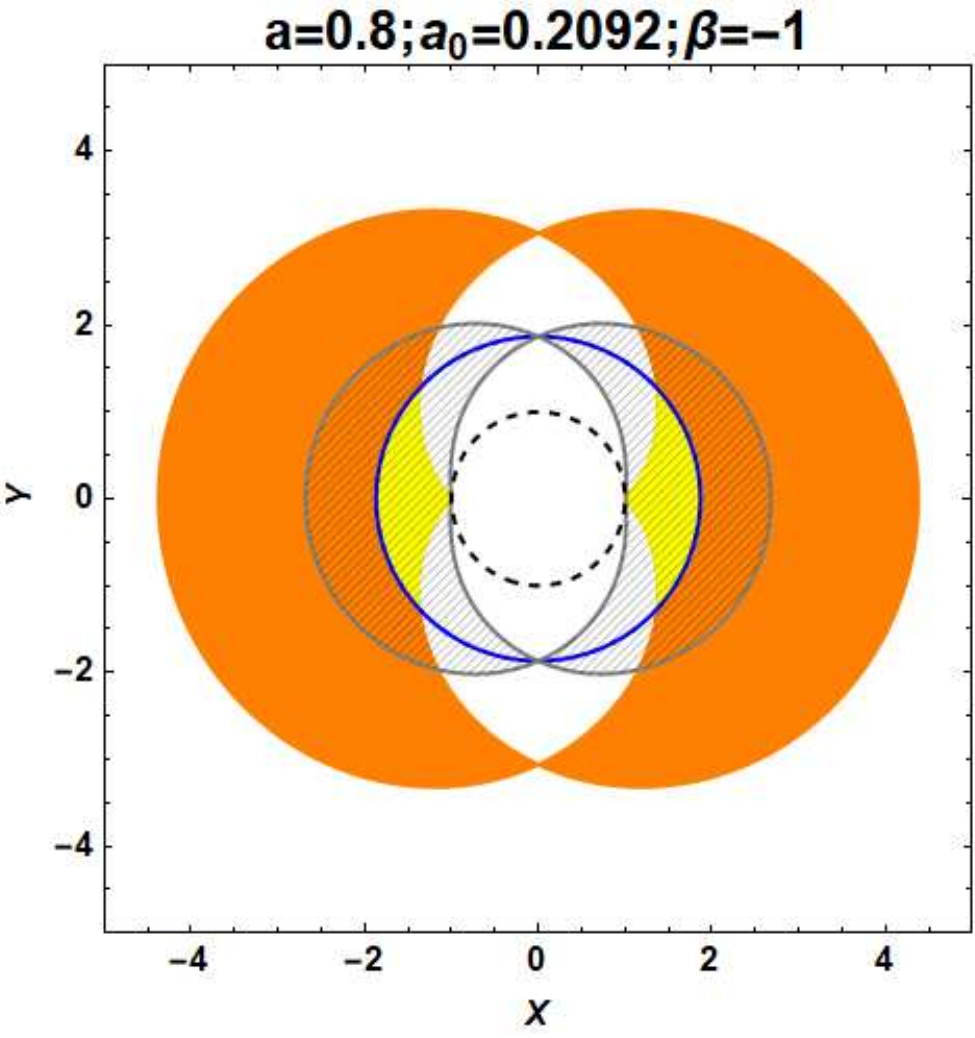}
  \hfill
\includegraphics[width=0.225\textwidth]{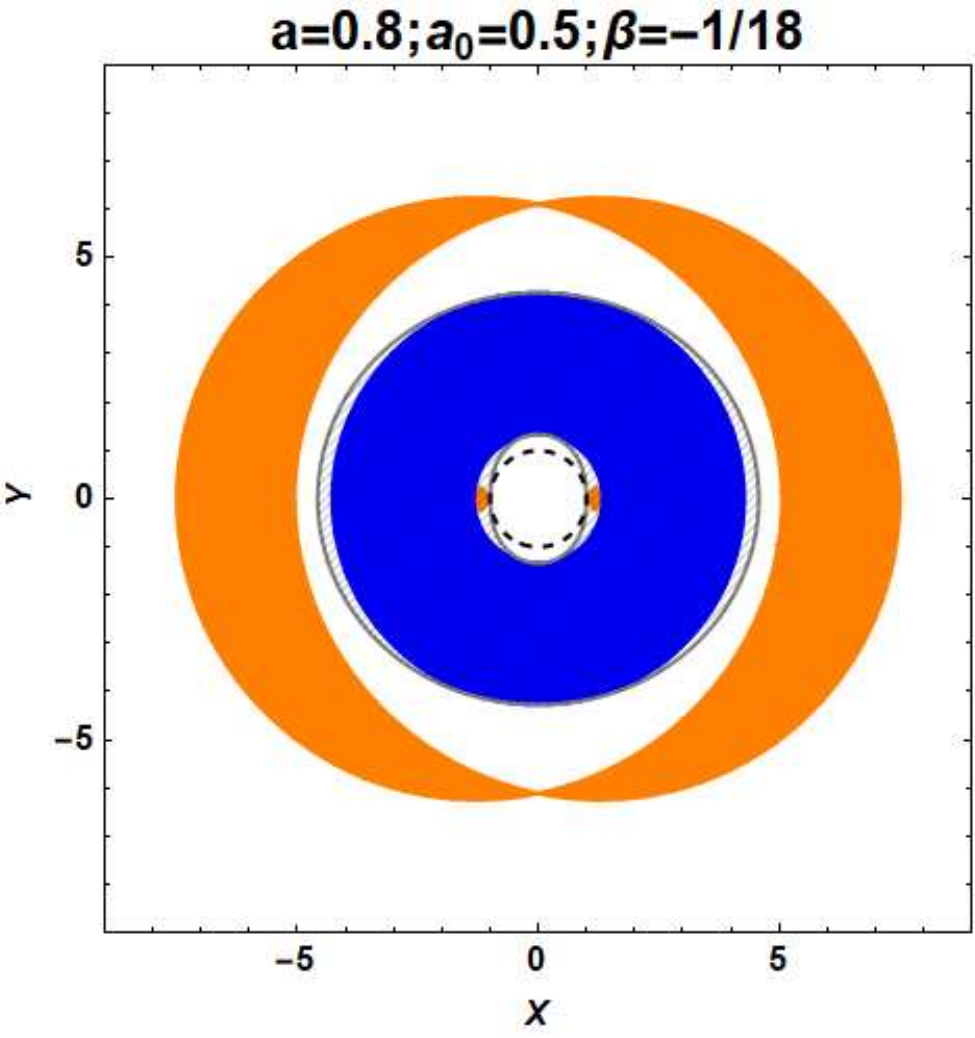}
\includegraphics[width=0.225\textwidth]{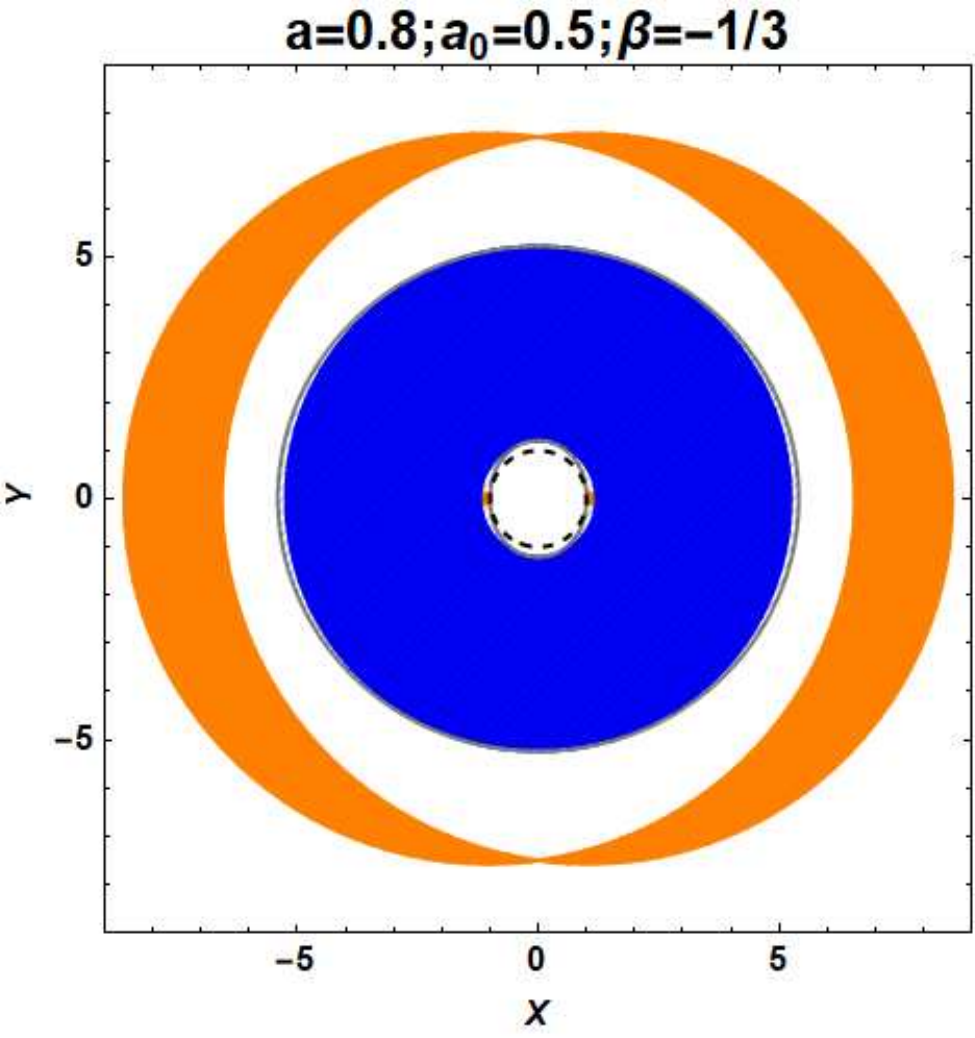}
 \includegraphics[width=0.225\textwidth]{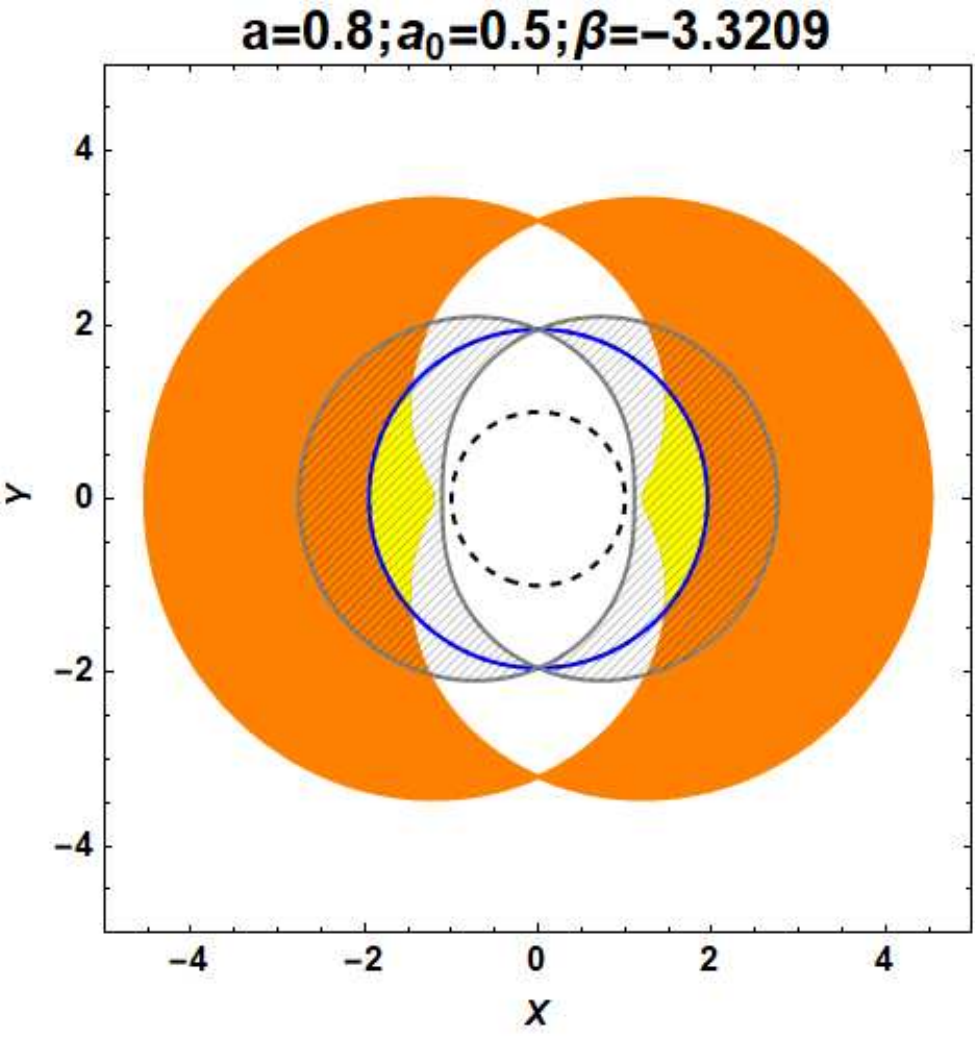}
  \includegraphics[width=0.225\textwidth]{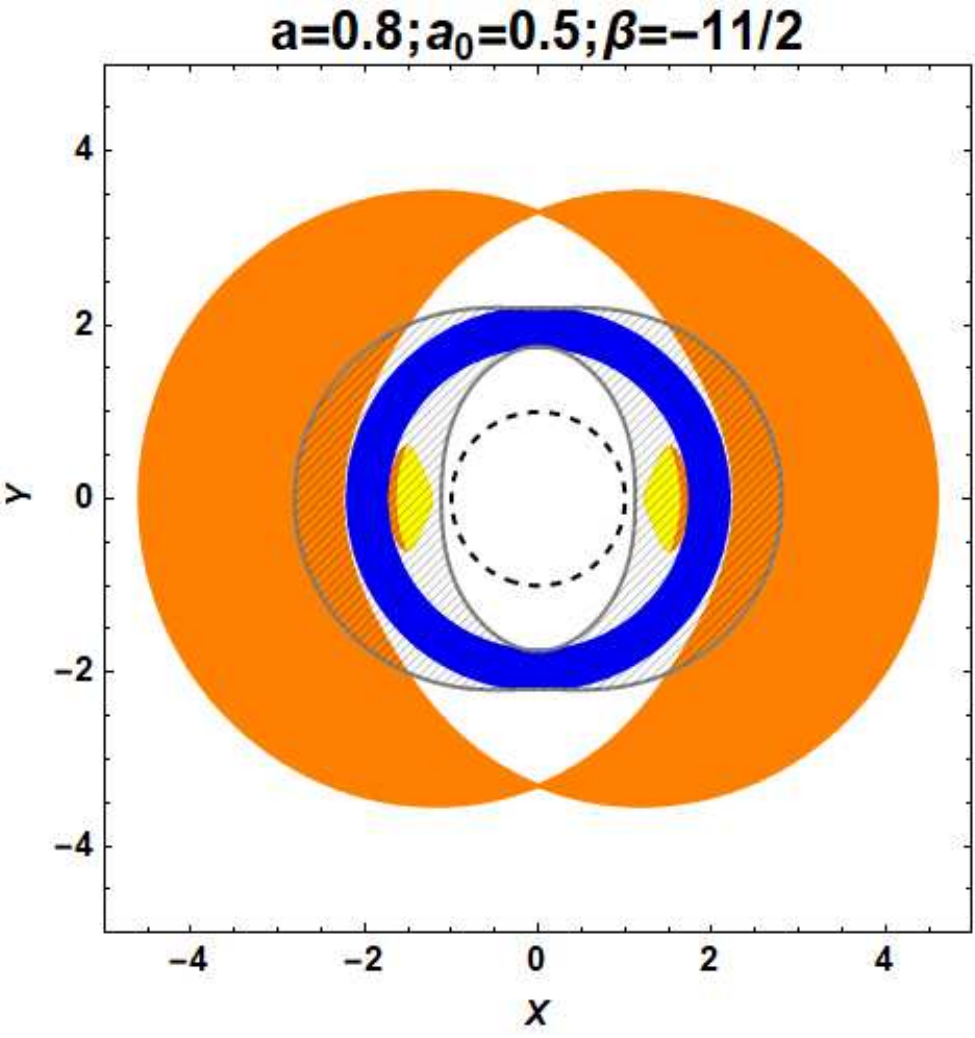}
  \caption{When $\beta<0$, the photon regions and ergosphere region of Kerr black hole surrounded by a cloud of strings in Rastall gravity.}
       \label{xc}
 \end{figure*}

Putting the above obtained equations into Eq.~(\ref{theta}), we can get the condition for the existence of the photon region,
\begin{equation}
(4r \Delta_r-\Sigma \Delta_r^{\prime})^2\le 16a^2r^2 \Delta_r\sin \theta ^2.
 \end{equation}

For radial perturbations, the spherical null geodesic at $r=r_p$ can be unstable or stable in Boyer-Lindquist coordinates, depending on the sign of the $R^{\prime \prime}(r)$. With the help of Eq.~(\ref{r}) and Eq.~(\ref{kele}), one can write
\begin{equation}
\frac{R^{\prime \prime}(r)}{8 E^{2}}\left(\Delta_{r}^{\prime}\right)^{2}=2 r \Delta_{r} \Delta_{r}^{\prime}+r^{2}\left(\Delta_{r}^{\prime}\right)^{2}-2 r^{2} \Delta_{r} \Delta_{r}^{\prime \prime}.
\end{equation}

When $R^{\prime \prime}(r)>0$, it means that the spherical null geodesic is unstable, while it implies that the spherical null geodesic is stable when $R^{\prime \prime}(r)<0$. In order to draw the photon region more clearly and completely, according to Refs.~\cite{Meng:2022kjs,Grenzebach:2014fha,Grenzebach:2015oea}, we adopt two different scales to cover all situations and exaggerate the external parts. Specifically, one employs $m\exp{(r/m)}$ in the area of $r<0$ and $r+m$ for the area of $r>0$. Furthermore, we use a black dotted line to represent the throat with $r=0$.

In Figs. \ref{dc} and \ref{xc}, we draw the photon region of the black hole, where orange represents unstable photon orbits and yellow represents stable photon orbits. The boundary of the blue area ($\Delta_{r}\le0$) represents the horizons of the black hole. The gray area ($g_{tt}>0$) corresponds to the ergoregion, where the timelike killing vector becomes spacelike in this region. In the pictures, we can observe that there are unstable photon orbits in an exterior photon region which is outside the event horizon. Sometimes there exist unstable and stable photon orbits in the interior photon region which is inside the Cauchy horizon. In Fig. \ref{dc}, we first analyze the case of $\beta>0$. In the first line, we fix $a=0.85$, $\beta=\frac{11}{2}$, and observe the impact of $a_0$ on the ergosphere region and photon region. As $a_0$ increases, the exterior photon region increases, while the interior photon region first decreases and then increases. And the ergosphere region increases for an increase in $a_0$. Thus, the ergosphere region of Kerr black holes ($a_0=0$) is smaller than that of Kerr black holes surrounded by a cloud of strings in Rastall gravity. In the second line, we fix $a=0.85$, $a_0=0.8$ and discuss the effect of $\beta$ on the ergosphere and photon regions. It is obvious that the interior photon region gradually appears and the exterior photon region decreases slightly with the increase of $\beta$. Moreover, the ergosphere region increases with the increase in $\beta$.

Then we observe the case of $\beta<0$. In the first line, we fix $a=0.8$, $\beta=-1$, and observe the influence of $a_0$ on the ergosphere region and photon region. With the increase of $a_{0}$, the exterior photon region increases, and the interior photon region and ergosphere region enlarge. In the second line, we fix $a=0.8$, $a_0=0.5$ and find the influence of $\beta$ on the ergosphere region and photon region. The change of photon and ergosphere regions with the negative $\beta$ is complex in Fig. \ref{xc}.

\section{Black hole shadow} \label{bh shadow}
\begin{figure*}[ht!]
\centering
\includegraphics[width=0.3\textwidth]{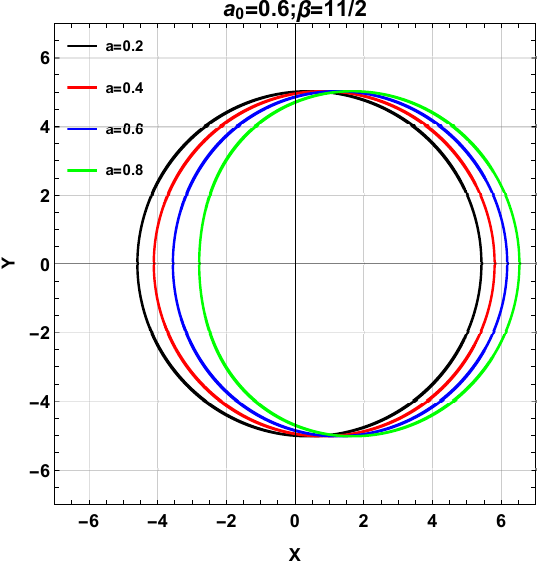}
\includegraphics[width=0.3\textwidth]{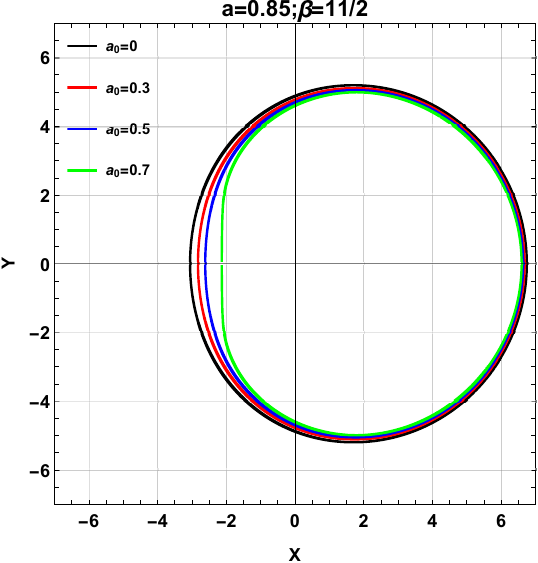}
\includegraphics[width=0.3\textwidth]{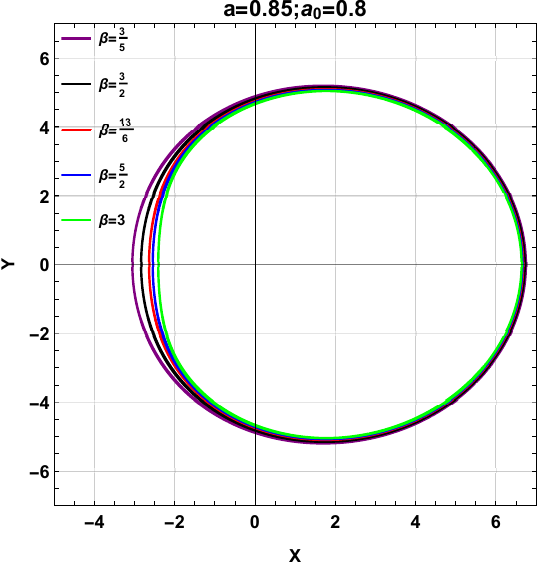}
   \caption{When $\beta>0$, the shadow geometry of Kerr black hole surrounded by a cloud of strings in Rastall gravity for different metirc parameters $a$, $a_{0}$ and $\beta$.}
       \label{ds}
 \end{figure*}
Considering that a black hole lies between the uniformly distributed light source and the observer, there are three possible trajectories for the photon geodesic when the photons from the light source arrive in the vicinity of $r_{+}$. Specifically, these three situations can be listed, namely, (i) swallowed by the black hole (capture orbits), (ii) fly to infinity (scatter orbits), and (iii) separate the two mentioned situations (unstable orbits). Observers in the domain of outer communication can only see scattered photons, while photons captured by the black hole form a 2D dark zone in the view of observers. This 2D dark zone is called the black hole shadow by observers. To visualize the black hole shadow, we consider an observer who is far away from the black hole so his neighborhood can be viewed as asymptotically flat. At the observer's sky, a closed curve will be displayed on the celestial coordinates (X,Y), resulting from the projection of the unstable photon orbits. Tracing a photon trajectory from the observer's position, a dot will be left on the image plane described by the following,

\begin{equation}
\begin{aligned}
&X=\lim_{r\to\infty}\left(-r^2\sin\theta\frac{d\phi}{dr}\bigg|_{\theta=\theta_0}\right), \\
&Y=\lim_{r\to\infty}\left(r^2\frac{d\theta}{dr}\bigg|_{\theta=\theta_0}\right).
\label{x}
\end{aligned}
\end{equation}

Here $\theta_{0}$ represents the inclination angle relative to the observer at infinity and $r$ indicates the distance from the black hole to the observer. Furthermore, Eq.~(\ref{x}) can be simplified to the following form by using the geodesic equations,
\begin{equation}
\begin{aligned}
X&=-\xi(r_p)\csc\theta_0,\\Y^2&=a^2\cos^2\theta_0+\eta(r_p)-\xi(r_p)^2\cot^2\theta_0,
\end{aligned}
\end{equation}
where $\xi(r_p)=L_E|_{r_p}\mathrm{~and~}\eta(r_p)=K_E-(L_E-a)^2|_{r_p}$. For the observer at the equatorial plane ($\theta_{0}=\pi/2$), it can be reduced to
\begin{equation}
\begin{aligned}
X&=-\xi(r_p),\\Y&=\pm\sqrt{\eta(r_p)}.
\end{aligned}
\end{equation}

\begin{figure*}[ht!]
\centering
 \includegraphics[width=0.3\textwidth]{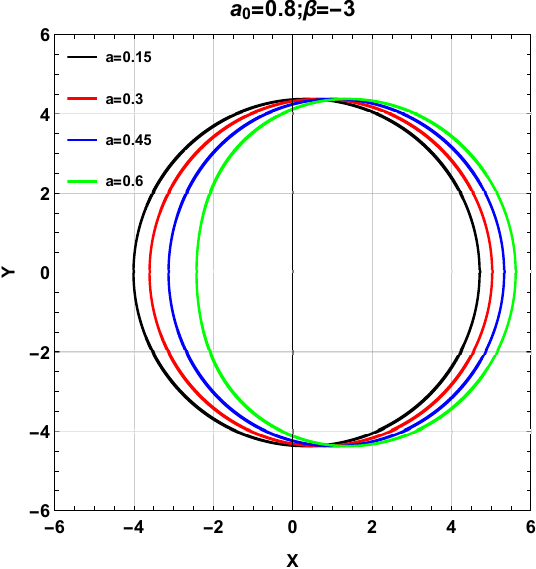}
\includegraphics[width=0.3\textwidth]{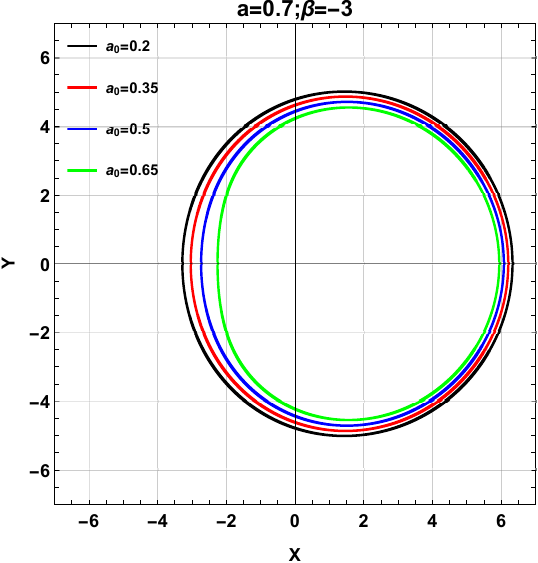}
\includegraphics[width=0.31\textwidth]{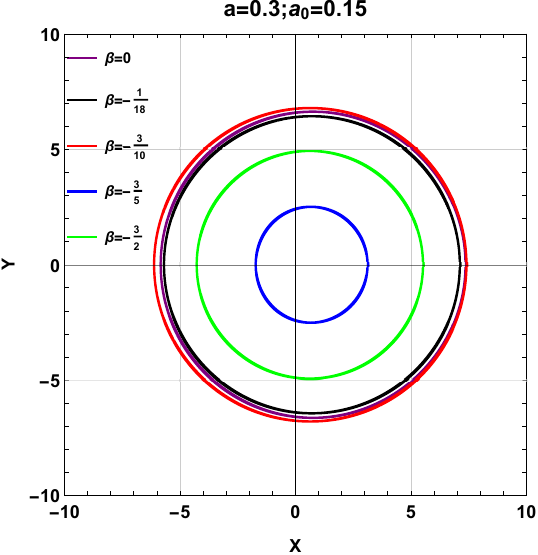}
    \caption{When $\beta<0$, the shadow geometry of Kerr black hole surrounded by a cloud of strings in Rastall gravity for different metirc parameters $a$, $a_{0}$ and $\beta$.}
       \label{xs}
 \end{figure*}

The region enclosed within the closed contours of $X$ and $Y$ in the above equation describes the shadow of the Kerr black hole surrounded by a cloud of strings under Rastall gravity. We show the black hole shadow varying with different parameters in Fig. \ref{ds} ($\beta>0$) and Fig. \ref{xs} ($\beta<0$).

In Fig. \ref{ds}, the left panel shows that the increase of $a$ leads to greater deformation, which depicts the asymmetry of the black hole shadow about the coordinate axis $x=0$ for $a_0=0.6$, $\beta=\frac{11}{2}$. In the middle panel, we show the comparison of the shadows of the Kerr black hole surrounded by a cloud of strings in Rastall gravity and the Kerr black hole. And it is implied that the shadow size of the Kerr black hole surrounded by a cloud of strings in Rastall gravity is smaller than the Kerr case ($a_0$ = 0). The increase of parameter $a_{0}$ is discovered to keep an incremental influence on the deformation of black hole shadow and decreases its size, for fixed $a=0.85$, $\beta=\frac{11}{2}$. In the right panel, we observe how the black hole shadow changes with varying $\beta$ when $a=0.85$, $a_0=0.8$. With the increase of $\beta$, it is obvious that the shadow size decreases and the deformation increases.

Additionally, we analyze the black hole shadow for different parameters $a$, $a_0$ and $\beta$ for $\beta<0$ in Fig. \ref{xs}. In the left panel, the deformation of the black hole shadow increases with the increase of the spin $a$ for fixed $a_0=0.8$, $\beta=-3$. In the middle panel, the parameter $a_0$ has a decremental effect on the size of black hole shadow and an incremental influence on the deformation for fixed $a=0.7$, $\beta=-3$. In the right panel, it shows the black hole shadow varies with parameter $\beta$ for $a=0.3$, $a_{0}=0.15$. With increasing parameter $\beta$, the size of the black hole shadow initially decreases, then increases, followed by another decrease, and finally increases again. And the size of the Kerr black hole surrounded by a cloud of strings in GR ($a_0=0$) and the size of the Kerr black hole surrounded by a cloud of strings in Rastall gravity cannot be directly judged.

\section{Shadow observables and parameter estimation} \label{observables}
From the previous section, we know that the size and shape of the photon ring are related to the parameters associated with the black hole. In other words, we can infer nature about black hole from the image of the black hole shadow, which is caused by gravitational field near the black hole \cite{Virbhadra:1999nm,Virbhadra:2002ju}. Recently, the EHT collaboration gained valuable data by observing the black hole in M87, which is an elliptical galaxy. These observations imply that the image of \text{M87*} is relevant to the Kerr black hole in GR. However, Refs.~\cite{EventHorizonTelescope:2019dse,EventHorizonTelescope:2019uob,EventHorizonTelescope:2019jan,EventHorizonTelescope:2019ths,EventHorizonTelescope:2019pgp,EventHorizonTelescope:2019ggy} did not consider the black hole in MoG. For a rotating black hole in MoG, its shadow will also be distorted due to the existence of additional parameters. This promotes us to estimate the parameters of Kerr black holes surrounded by a cloud of strings under Rastall gravity. Here we assume that the observer is on the equatorial plane, that is, the inclination angle $\theta=\frac{\pi}{2}$.

For the sake of extracting some information about black holes from the complex black hole shadow image, Hioki and Maeda \cite{Hioki:2009na} proposed to use the two characteristic parameters $R_S$ and $\delta _S$ for numerical estimation. Here $R_S$ and $\delta _S$ approximately describe the radius and distortion parameter of the black hole shadow, respectively. Regarding their calculation, we can first define a reference circle, whose top, bottom, right, and left coordinates are $(X_t,Y_t), (X_b,Y_b), (X_r,0), (X_{l }^{\prime},0)$, respectively. The leftmost coordinate of the black hole shadow is $(X_l,0)$. With these at hand, the two characterized observables can be defined by \cite{Hioki:2009na}
\begin{equation}
R_{S}=\frac{(X_{t}-X_{r})^{2}+Y_{t}^{2}}{2\mid X_{r}-X_{t}\mid},\quad\delta_{S}=\frac{\mid X_{l}-X_{l}^{\prime}\mid}{R_{S}}.
\end{equation}

\begin{figure*}
\centering
 \includegraphics[width=0.425\textwidth]{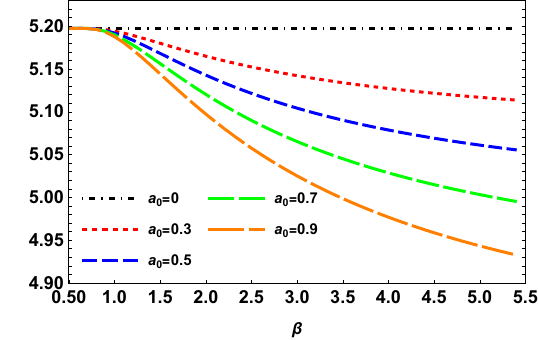}
\includegraphics[width=0.43\textwidth]{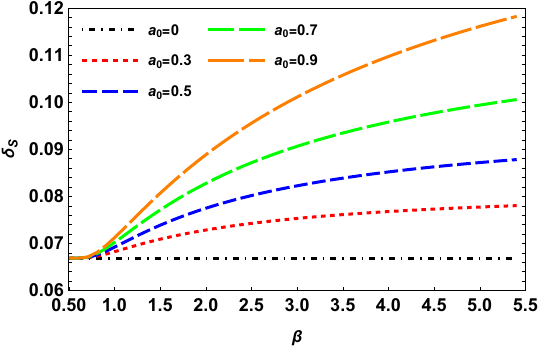}
 \includegraphics[width=0.42\textwidth]{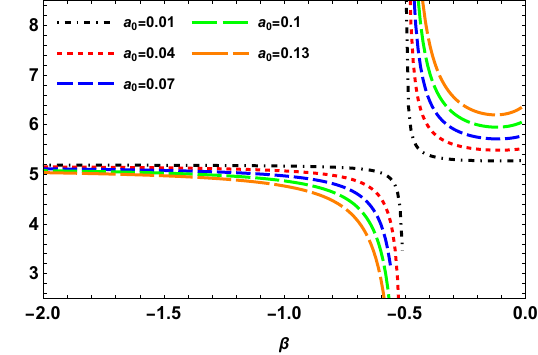}
\includegraphics[width=0.46\textwidth]{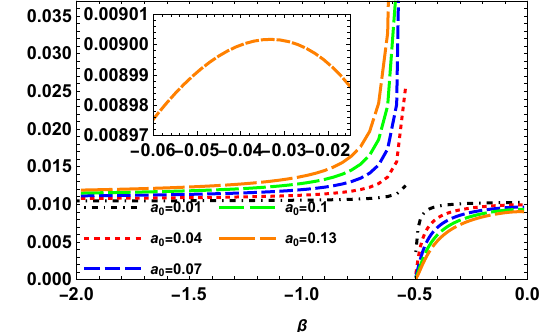}
    \caption{The shadow radius $R_{S}$ and distortion $\delta_{S}$ of the Kerr black hole surrounded by a clound of strings in Rastall gravity for $\beta>0$ when $a=0.7$ (the upper row) and $\beta<0$ when $a=0.3$ (the bottom row).}
       \label{rd}
 \end{figure*}

The upper panel of Fig. \ref{rd} shows the behaviour of observables $R_{S}$ and $\delta_S$ as a function of $\beta$ when $\beta>0$.
We find that the presence of $a_0$ decreases the shadow radius $R_{S}$ for the fixed x-axis, i.e., $\beta=constant$. Namely, the shadow radius of the Kerr black hole ($a_{0}=0$) is larger than that of the Kerr black hole surrounded by a cloud of strings in Rastall gravity. And it can be clearly seen that the radius $R_{S}$ decreases with the increase of $\beta$ and $a_{0}$. Then, the influence of metric parameters $a_{0}$ and $\beta$ on distortion parameter $\delta_{S}$ is explored. The presence of $a_{0}$ increases the distortion of the black hole. Additionally, the distortion of the black hole increases with the increase of $a_0$ and $\beta$.

For the lower panel of Fig. \ref{rd}, it depicts the behaviour of observable $R_{S}$ and $\delta_S$ as a function of $\beta$ for $\beta<0$. The denominators of $\lambda_1$ and $\lambda_2$ in metric being nonzero requires that $\beta \neq \pm \frac{1}{2}$. So there is a discontinuity point $\beta =-\frac{1}{2}$, and the entire negative axis is divided into two regions.

When we fix the x-axis and investigate the influence of $a_0$ on the black hole shadow radius, we find that the black hole shadow radius decreases with an increase in $a_{0}$ when $\beta \in (-2, -\frac{1}{2})$. But in the area of ($-\frac{1}{2}$, 0), the black hole shadow radius increases for an increase in $a_0$. The black hole shadow radius first decreases slowly and then rapidly decreases as $\beta$ increases for $\beta \in (-2, -\frac{1}{2})$. In the region of ($-\frac{1}{2},$ 0), although the black hole shadow radius first decreases and then increases, it is larger than the value in the region of ($-2$, $-\frac{1}{2}$). Moreover, the shadow radius of the black hole in region of ($-2$, $-\frac{1}{2}$) is smaller than that of the Kerr black hole surrounded by a cloud of strings in GR.

When we hold the x-axis constant and observe the influence of $a_0$ on the black hole distortion, we find that in the region of ($-2$, $-\frac{1}{2}$), the increase of $a_0$ leads to an increase in black hole distortion, while it decreases in the region of ($-\frac{1}{2}$, 0). Besides, the distortion of black hole slowly increases and then sharply rises as $\beta$ increases in the region of ($-2$, $-\frac{1}{2}$). Even though the distortion of the black hole first increases and then decreases within ($-\frac{1}{2}$, 0), its value remains smaller than that in the range of ($-2$, $-\frac{1}{2}$). The distortion of the black hole for $\beta \in (-2, -\frac{1}{2})$ is greater than that of the Kerr black hole surrounded by a cloud of strings in GR. This can also be partially reflected by the image of shadow radius.

\begin{figure*}[ht!]
\centering
 \includegraphics[width=0.425\textwidth]{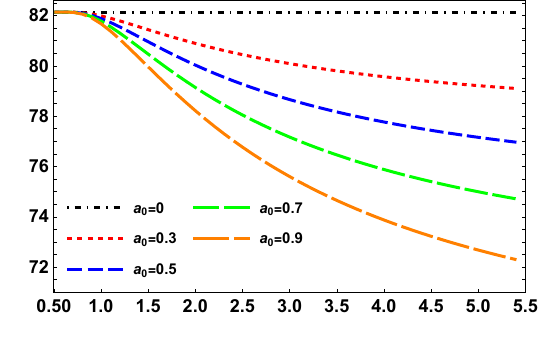}
  \includegraphics[width=0.442\textwidth]{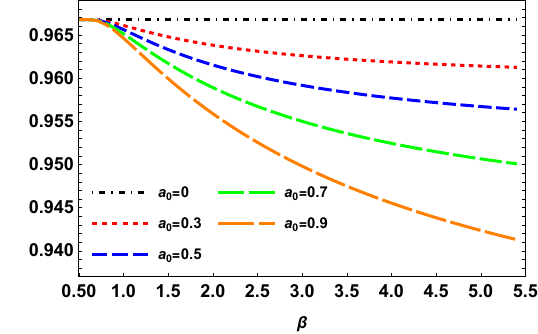}
 \includegraphics[width=0.435\textwidth]{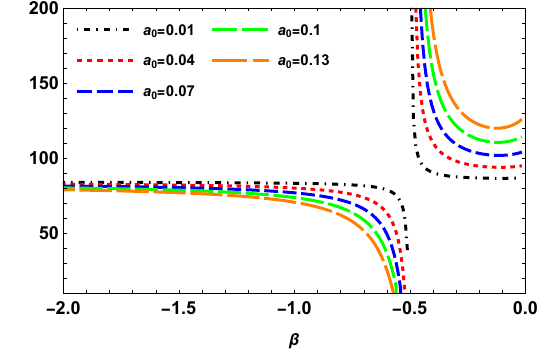}
\includegraphics[width=0.45\textwidth]{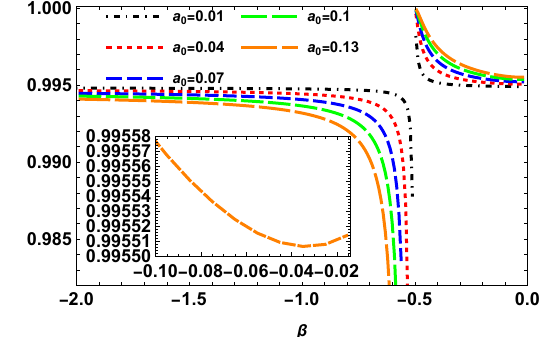}
    \caption{The shadow area $A$ and oblateness $D$ of the Kerr black hole surrounded by a clound of strings in Rastall gravity for $\beta>0$ when $a=0.7$ (the upper row) and $\beta<0$ when $a=0.3$ (the bottom row).}
       \label{add}
 \end{figure*}

In the view of $R_S$ and $\delta _S$ requiring a certain symmetry of the black hole shadow, they can only estimate the black hole parameters to some extent. So they cannot be applied well to extract the information about haphazard black holes shadows that appears in the MoG. Therefore, the coordinate-independent form of the shaded areas $A$ and oblateness $D$ are introduced \cite{Kumar:2018ple},
\begin{equation}
\begin{aligned}
A&=2\int Y(r_{p})dX(r_{p})=2\int_{r_{p}^{min}}^{r_{p}^{max}}\left(Y(r_{p})\frac{dX(r_{p})}{dr_{p}}\right)dr_{p},\\ D&=\frac{X_{r}-X_{l}}{Y_{t}-Y_{b}}.
\label{AD}
\end{aligned}
\end{equation}

\begin{figure*}[ht!]
\centering
 \includegraphics[width=0.42\textwidth]{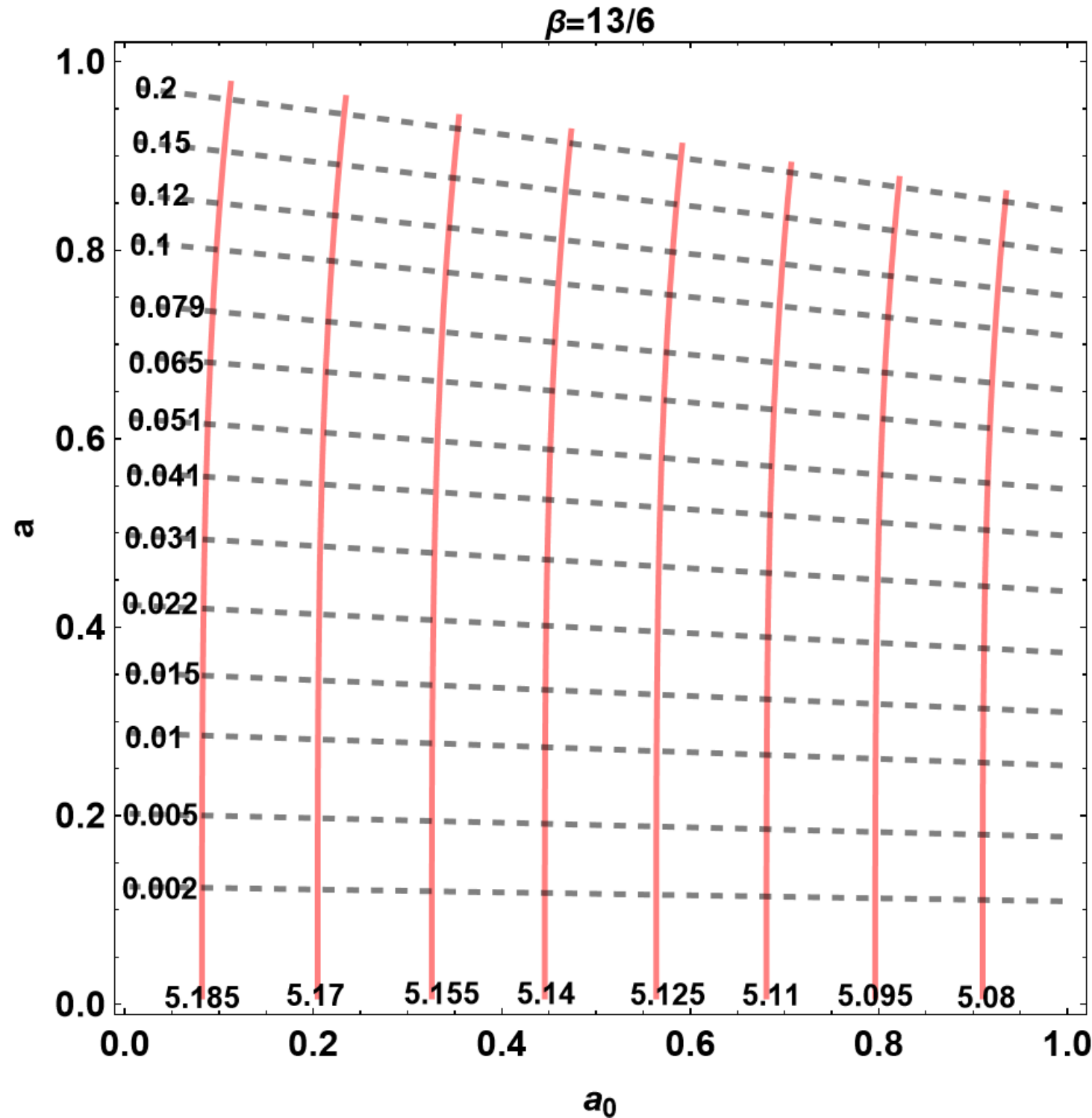}
\includegraphics[width=0.42\textwidth]{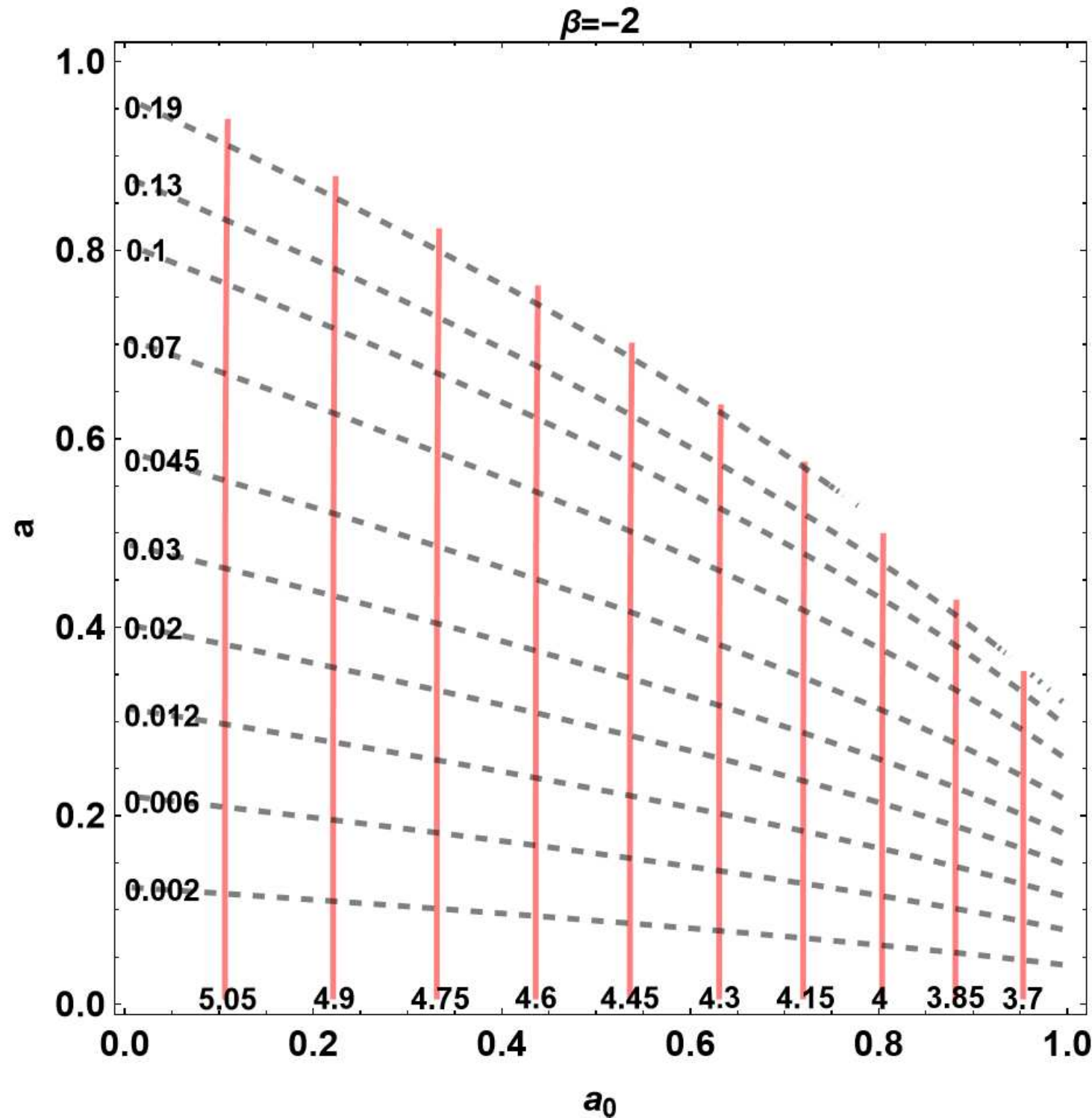}
 \includegraphics[width=0.42\textwidth]{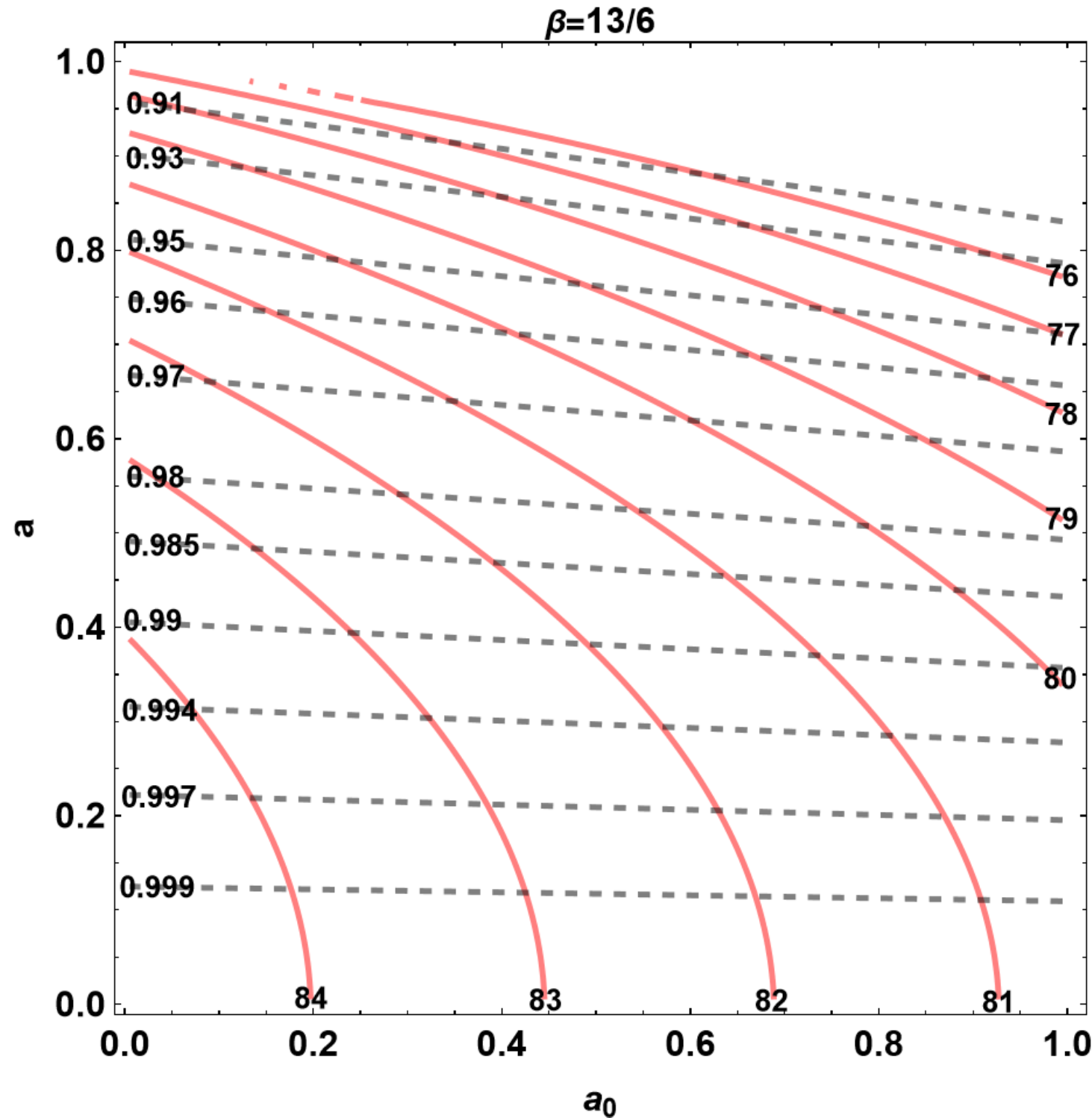}
\includegraphics[width=0.42\textwidth]{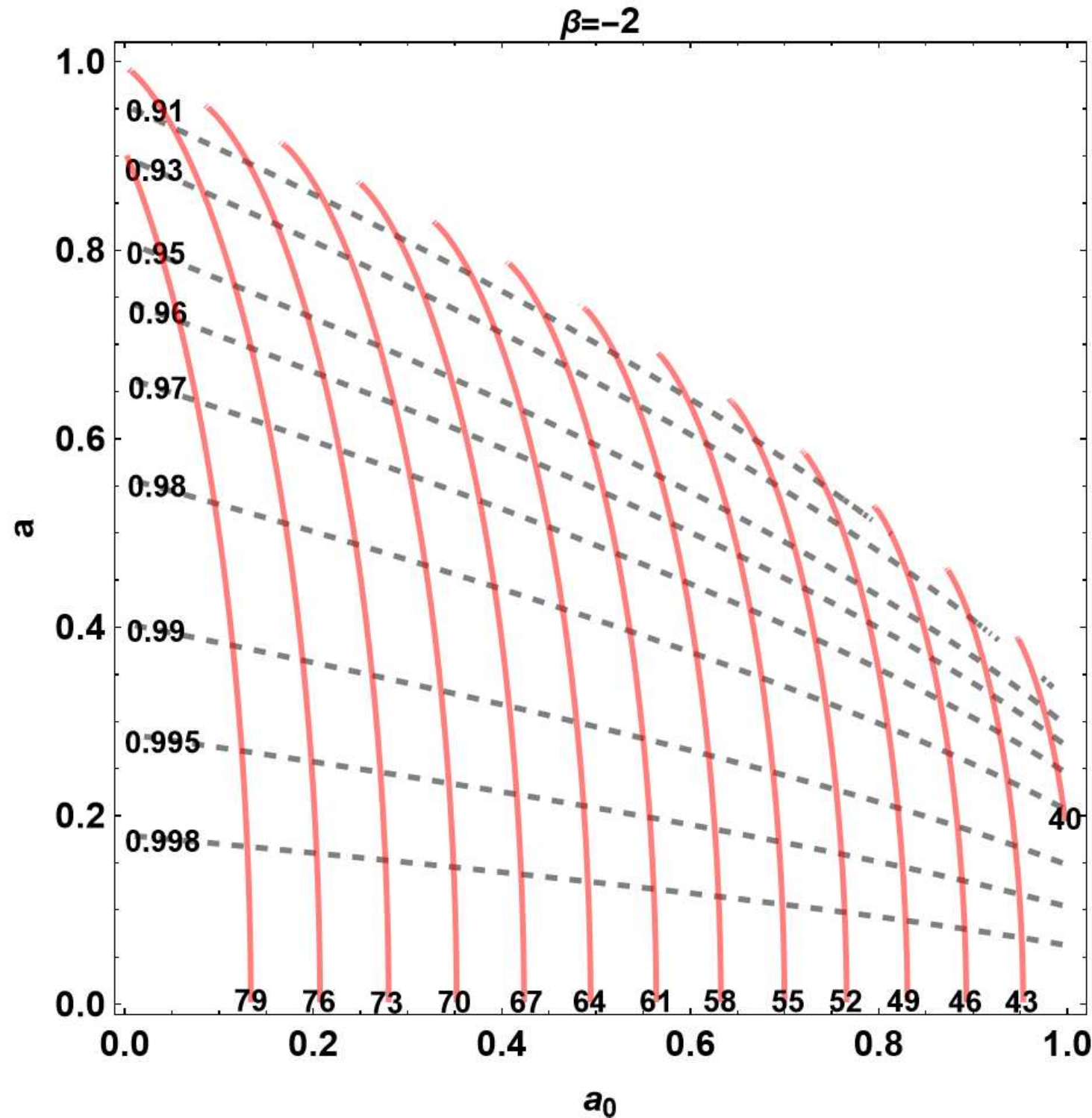}
    \caption{Upper row: the contour maps for shadow observables $R_{S}$ (red) and $\delta_{S}$ (black) in the parameter $(a, a_{0})$ of the Kerr black hole surrounded by a cloud of strings in Rastall gravity at inclinations for $\theta=\frac{\pi}{2}$. Bottom row: the contour maps for shadow observables $A$ (red) and $D$  (black) at inclinations for $\theta=\frac{\pi}{2}$.}
       \label{dg}
 \end{figure*}
 \begin{figure*}
\centering
 \includegraphics[width=0.43\textwidth]{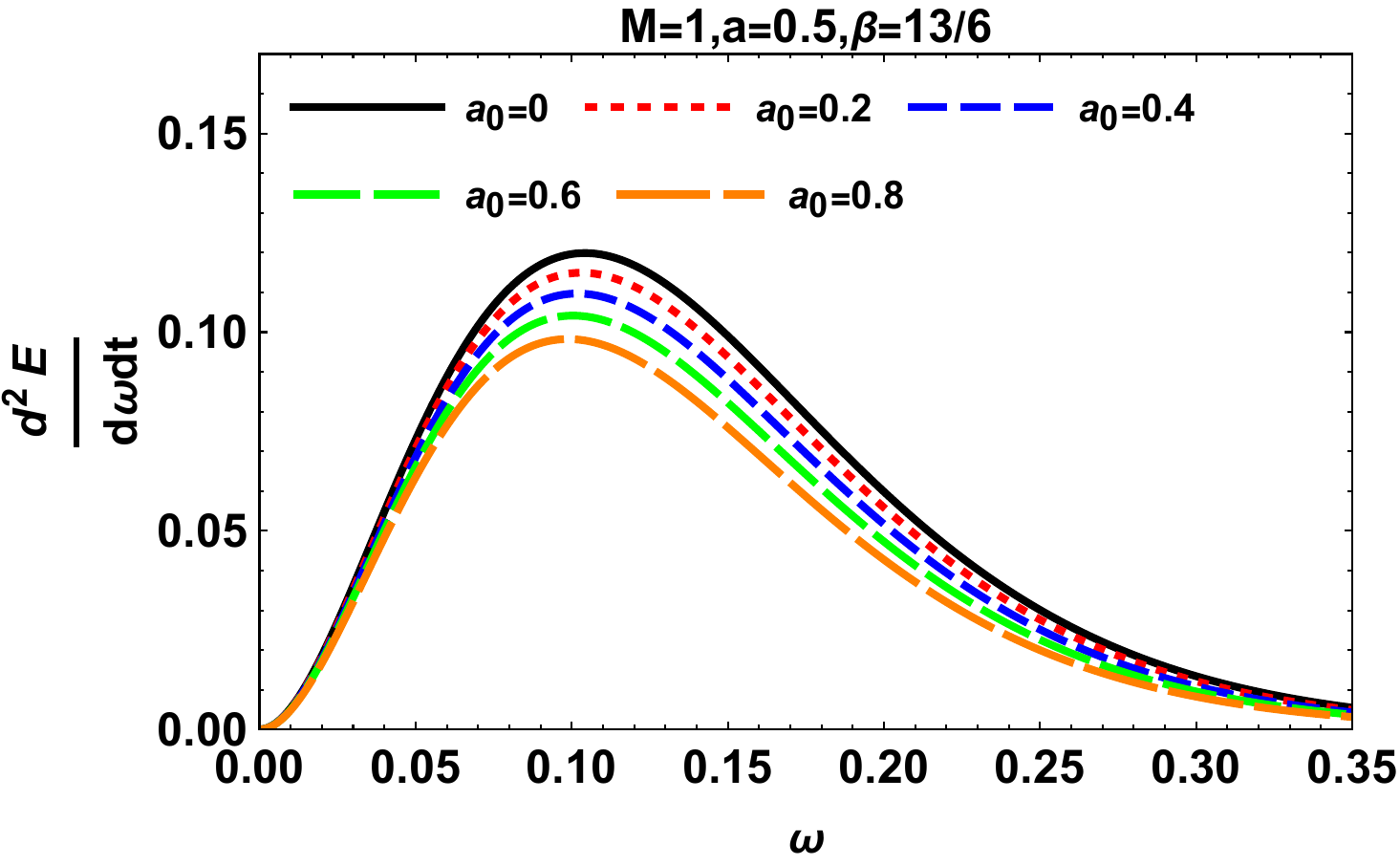}
\includegraphics[width=0.415\textwidth]{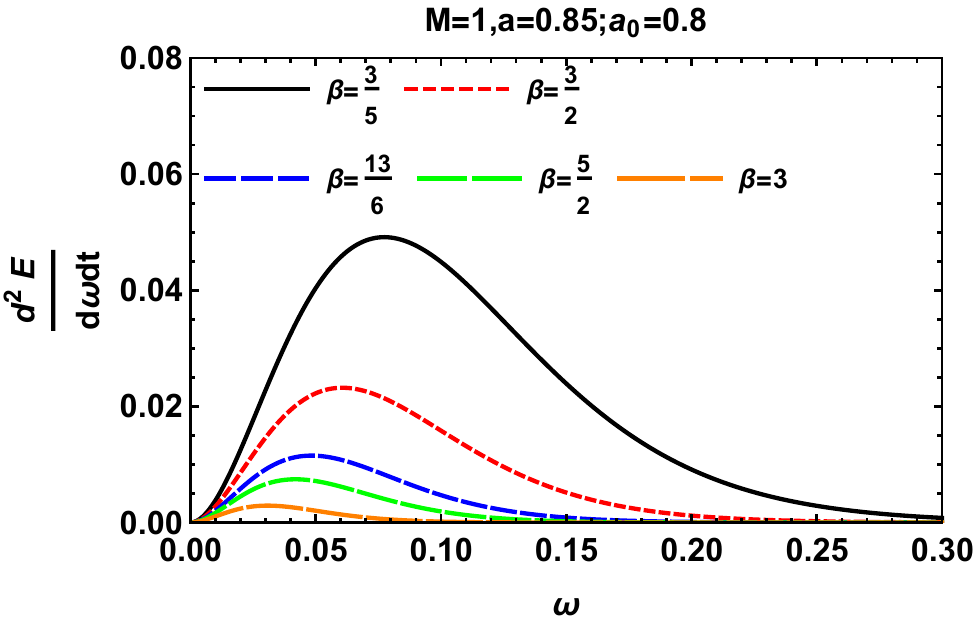}
 \includegraphics[width=0.43\textwidth]{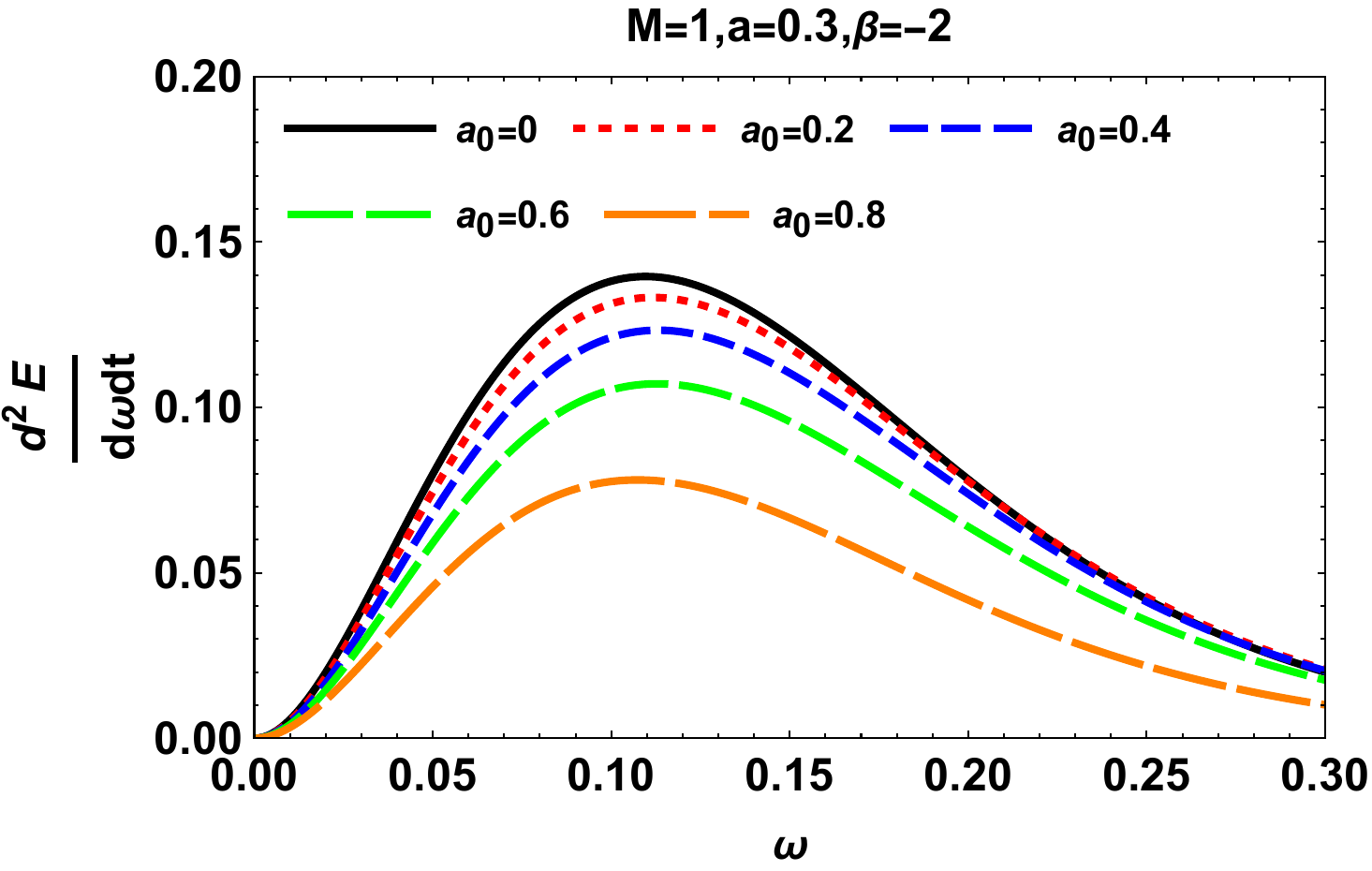}
\includegraphics[width=0.43\textwidth]{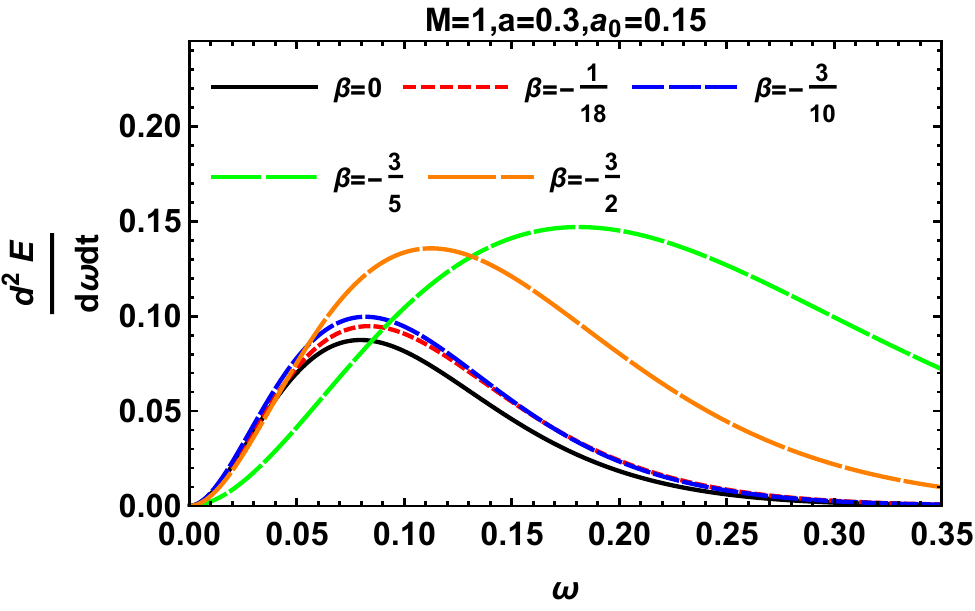}
    \caption{The energy emission rate of the Kerr black hole surrounded by a clound of strings in Rastall gravity for $\beta>0$ (the upper row) and $\beta<0$ (the bottom row).}
       \label{xe}
 \end{figure*}
\begin{figure*}[ht!]
\centering
 \includegraphics[width=0.43\textwidth]{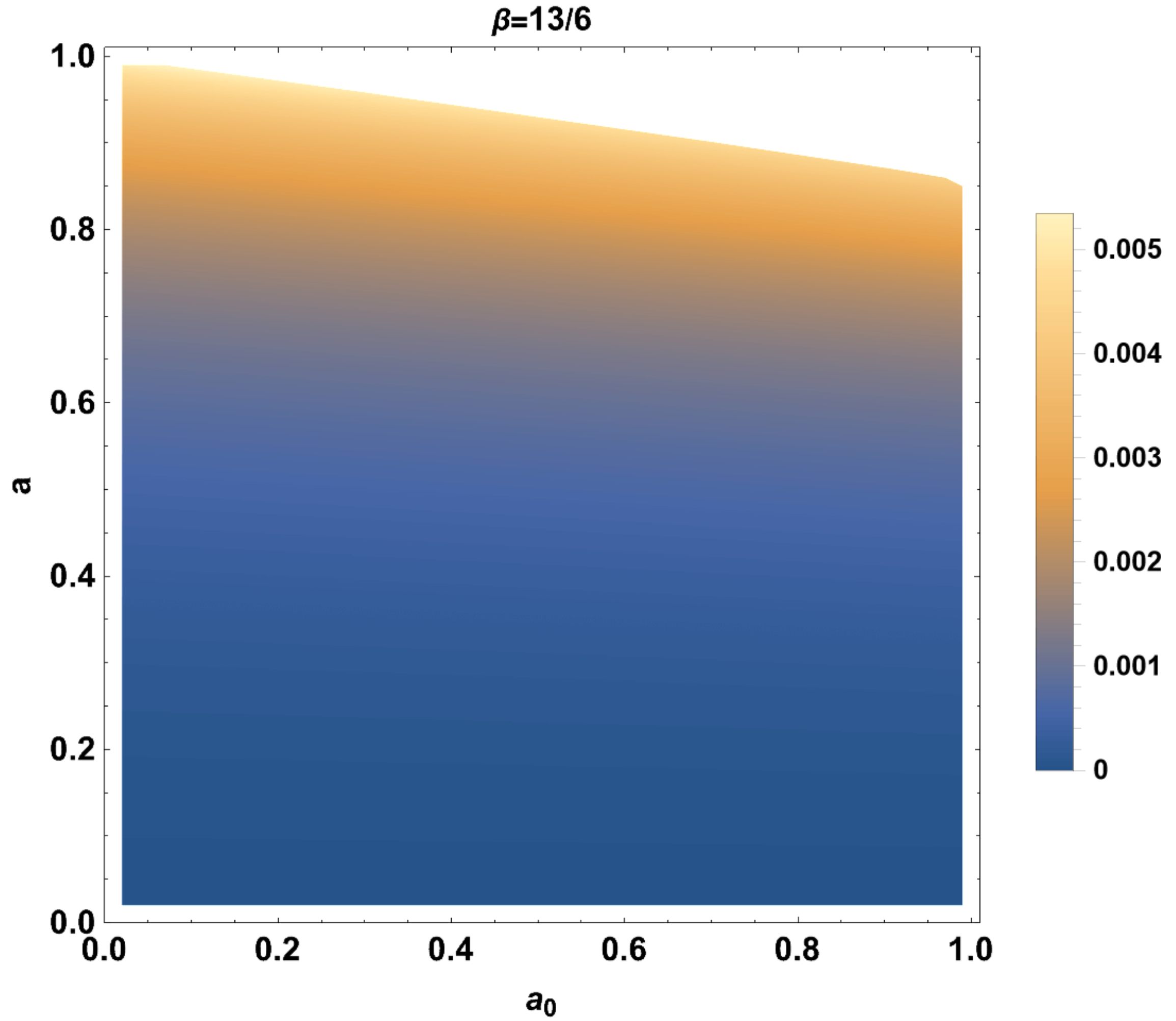}
\includegraphics[width=0.43\textwidth]{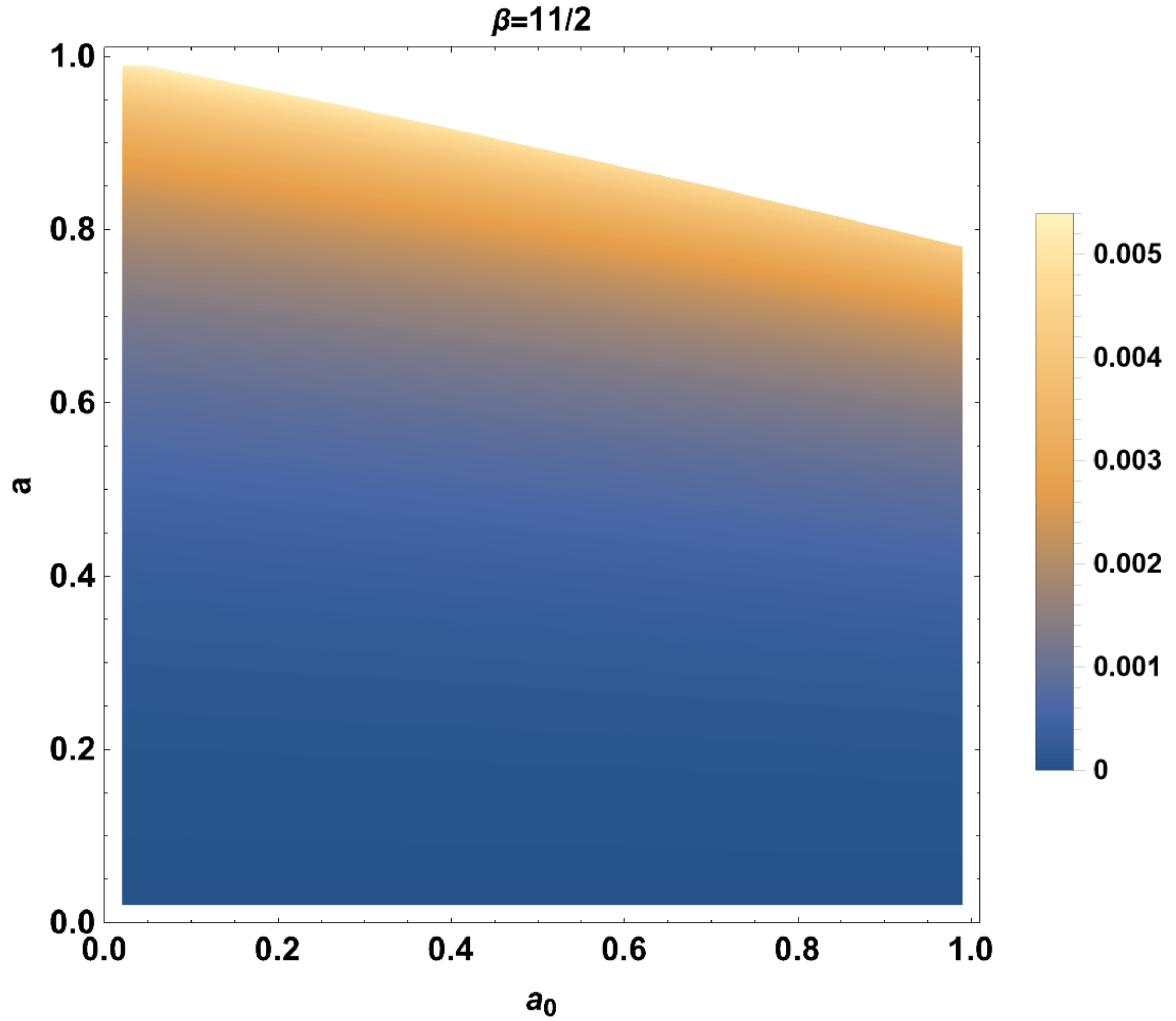}
\includegraphics[width=0.43\textwidth]{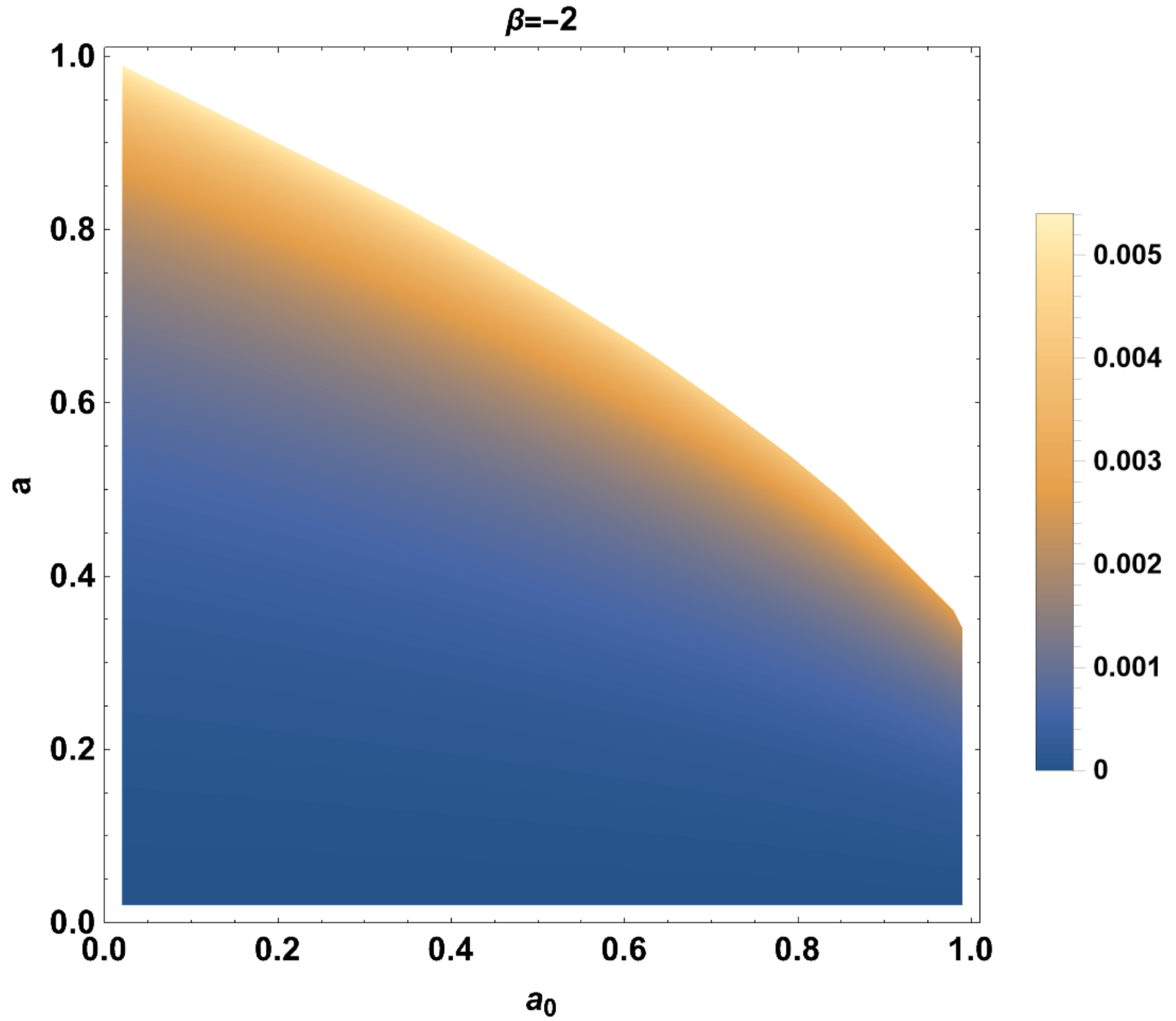}
\includegraphics[width=0.43\textwidth]{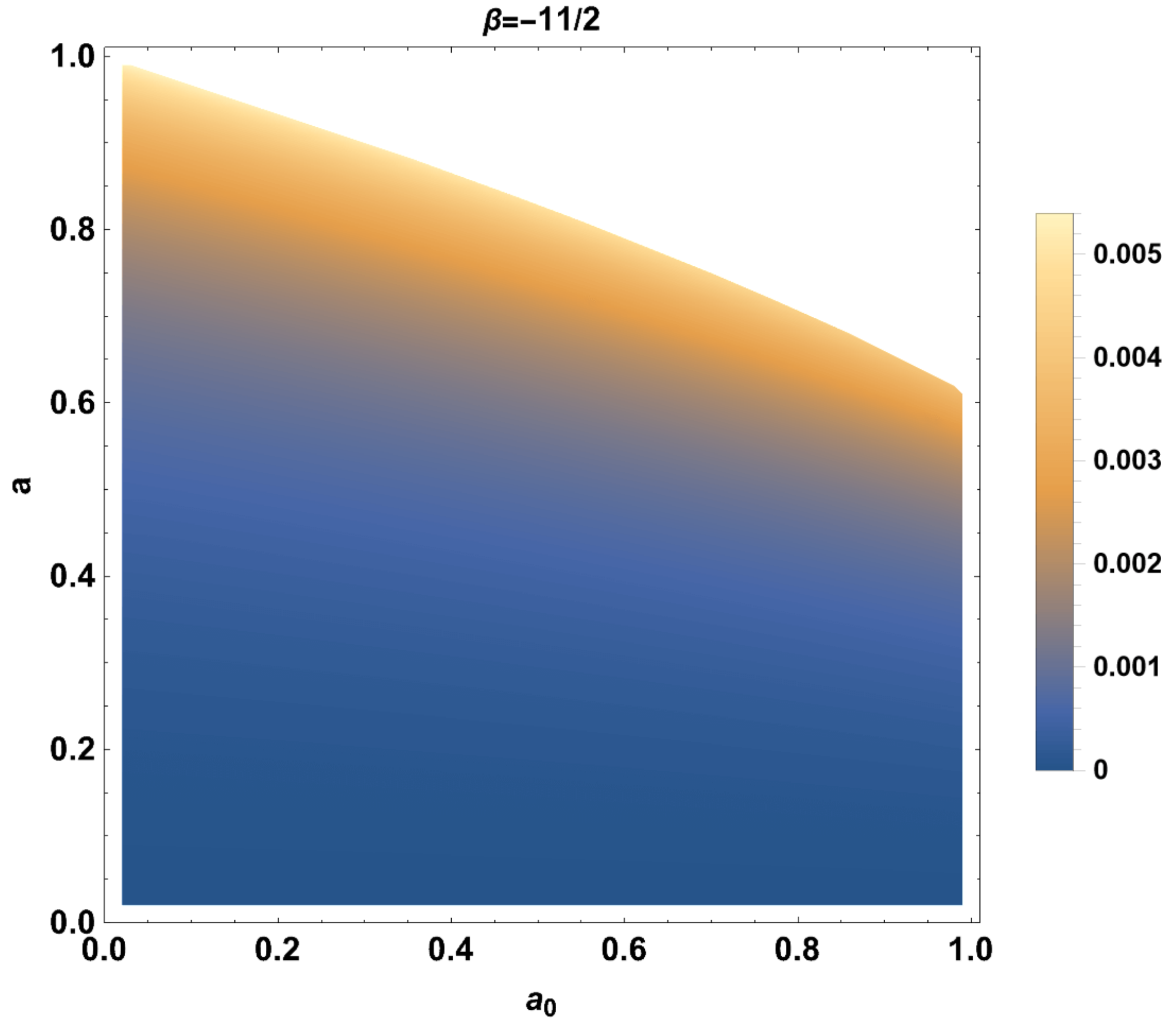}
    \caption{The density plots of  circularity deviation $\Delta C$ as function of $a$ and $a_{0}$ at inclinations for $\theta=17^{\circ}$.}
       \label{detalc}
 \end{figure*}
 \begin{figure*}[ht!]
\centering
 \includegraphics[width=0.43\textwidth]{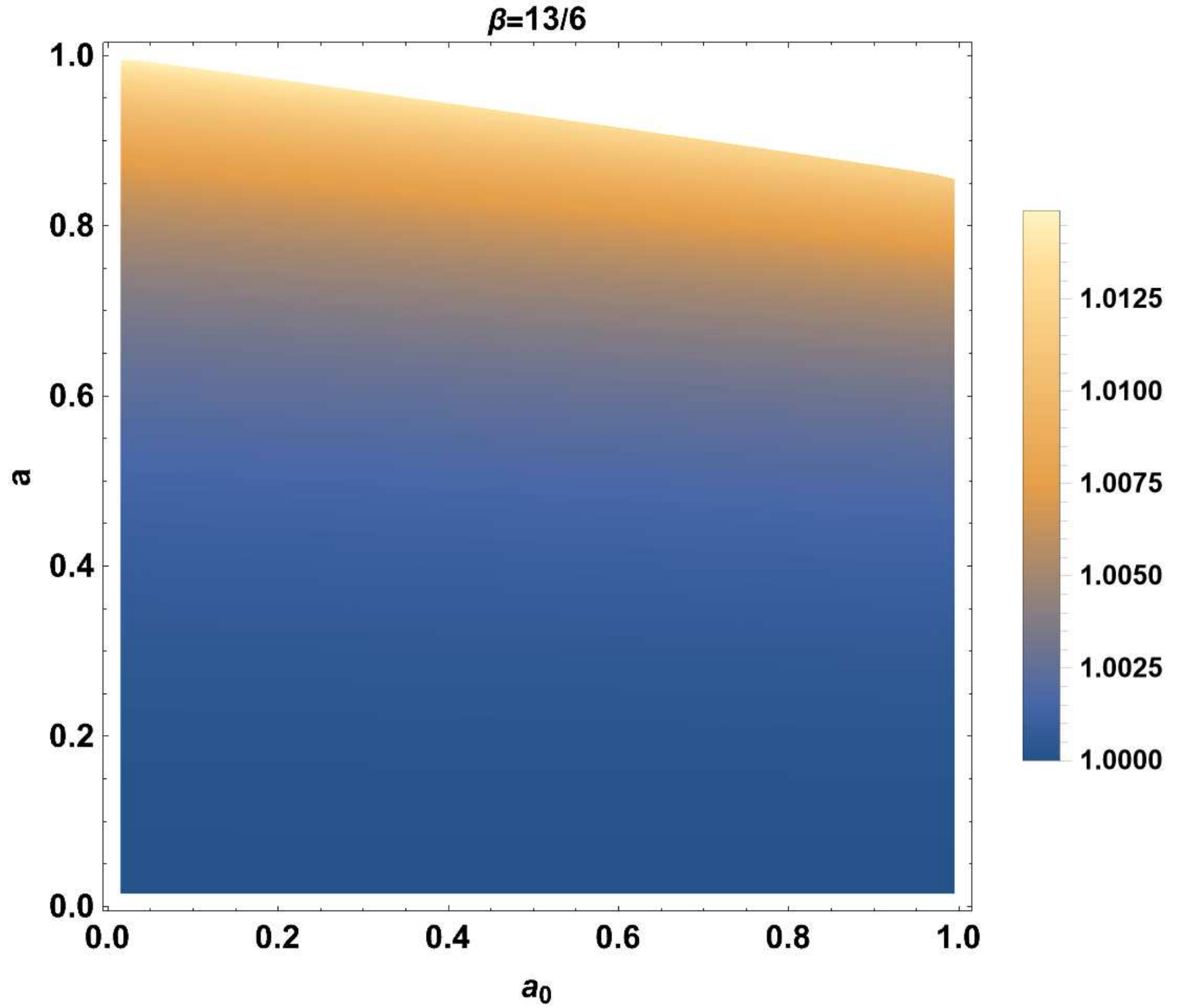}
\includegraphics[width=0.43\textwidth]{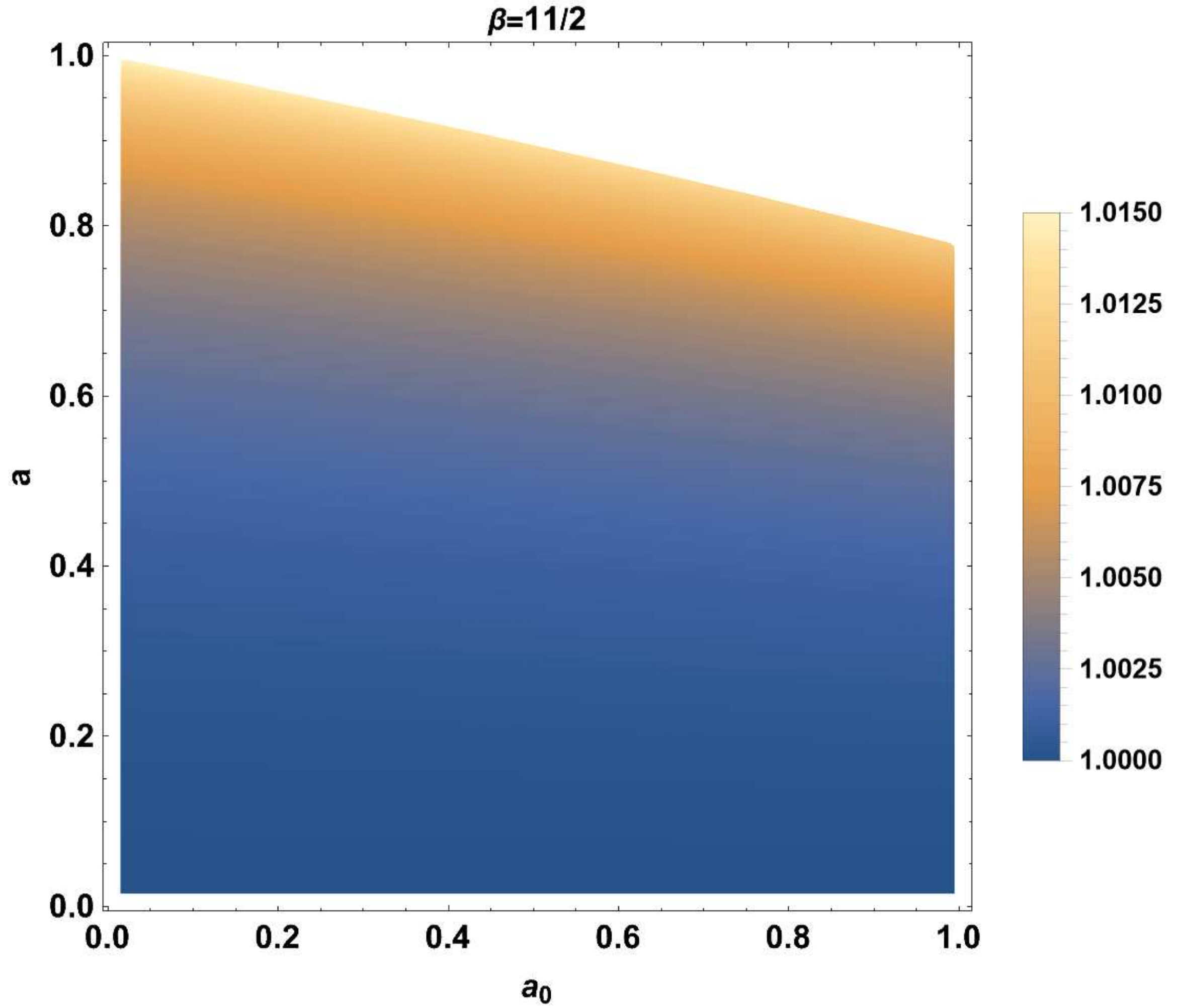}
 \includegraphics[width=0.43\textwidth]{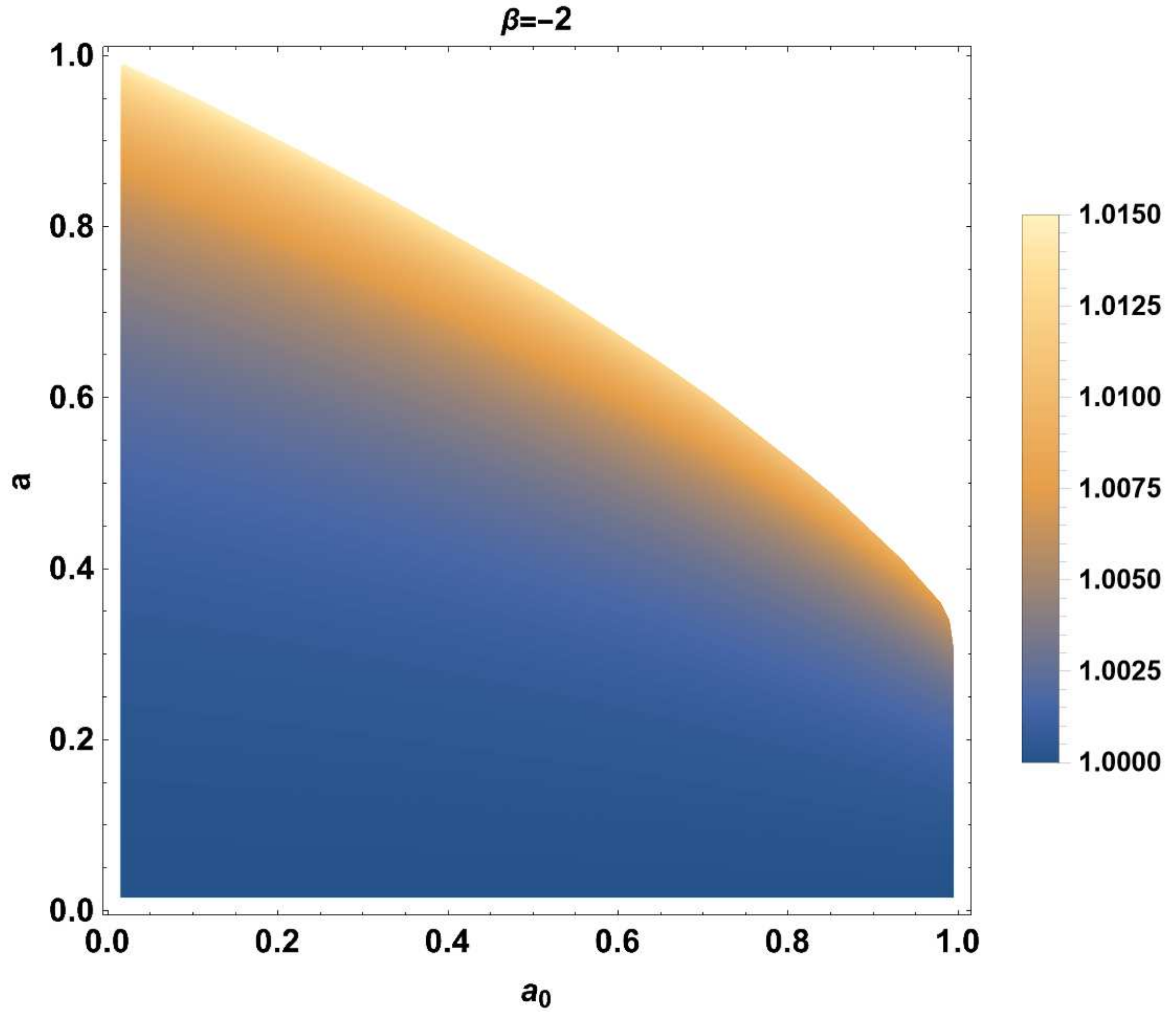}
\includegraphics[width=0.43\textwidth]{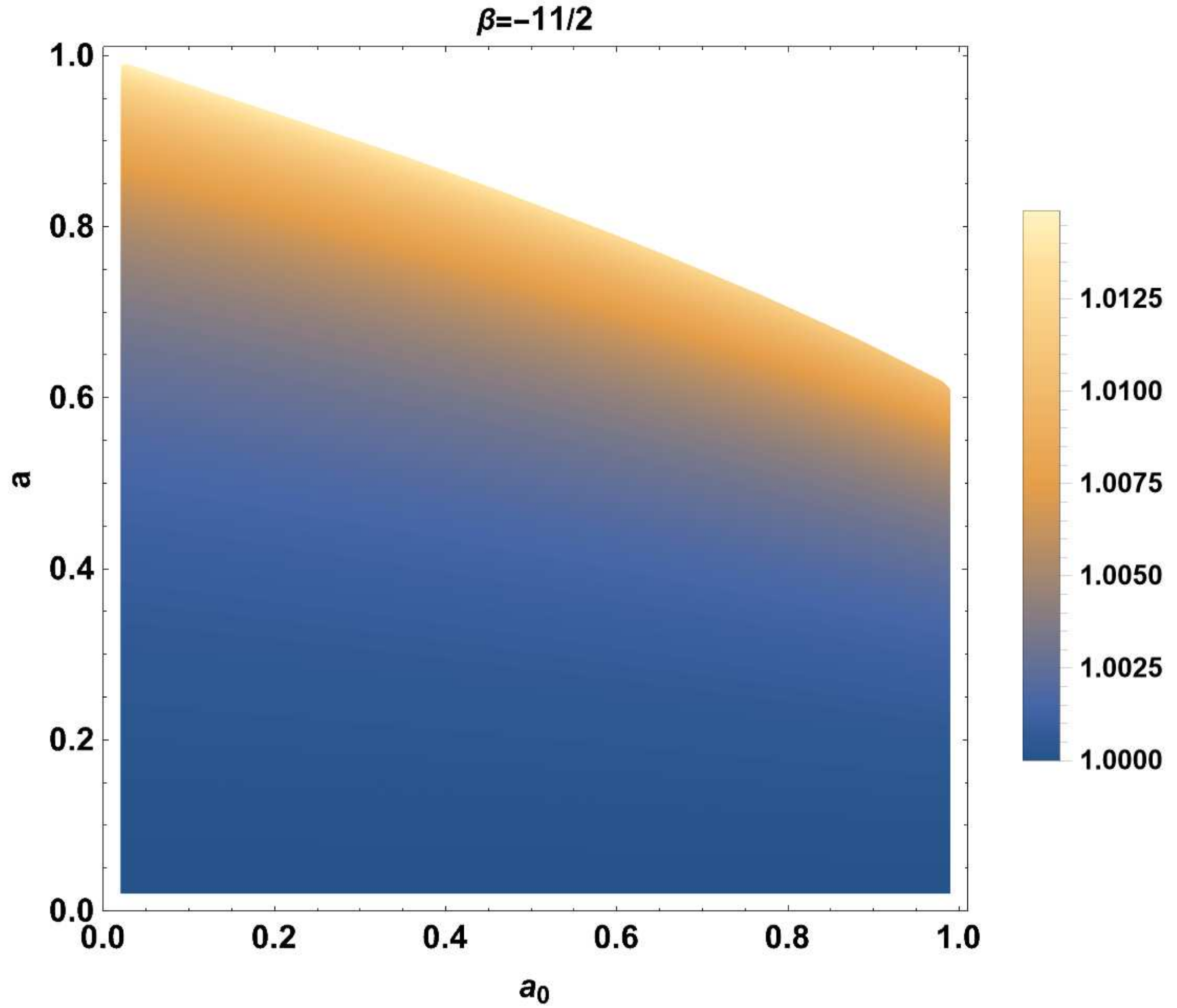}
    \caption{The density plots of axis ratio $D_{x}$ as function of $a$ and $a_{0}$ at inclinations for $\theta=17^{\circ}$.}
       \label{dx}
 \end{figure*}

Here $r_{p}^{min}$ and $r_{p}^{max}$ are the radii of prograde and retrograde orbits obtained by $\eta(r_p)=0$. For observers on the equatorial plane, $D=1$ and $D=\sqrt{3}/2$ correspond to the conditions of the Schwarzschild black hole and the extremal Kerr black holes, respectively. $\sqrt{3}/2\le D<1$ corresponds to the situation of Kerr black hole.

Fig. \ref{add} shows the change of $A$ and $D$ with $\beta$. The upper row of the graph is the case when $\beta>0$. We can see that $A$ and $D$ both decrease with an increase in $\beta$. When we fix $\beta$ and observe the influence of $a_0$ on $A$ and $D$, it is obvious that $A$ and $D$ decrease with the increase of $a_0$. These correspond to the result that the shadows of the Kerr black hole surrounded by a cloud of strings in Rastall gravity are smaller and more distorted than a Kerr black hole which is plotted by the black dash-dot curve.

The lower row of Fig. \ref{add} shows the changes of $A$ and $D$ with $\beta$ when $\beta<0$. In the region of ($-2$, $-\frac{1}{2}$), both $A$ and $D$ perform a gradual decrease followed by a sharp decline as $\beta$ increases. In the region of ($-\frac{1}{2}$, 0), $A$ and $D$ first decrease and then increase for an increase in $\beta$. But the values are larger than those in the region of ($-2$, $-\frac{1}{2}$). When we fix $\beta$ and observe the two observables varying with $a_0$, we find that the values of $A$ and $D$ decrease with the increase of $a_0$ in the region of ($-2$, $-\frac{1}{2}$), while increase in the region of ($-\frac{1}{2}$, 0). The values of $A$ and $D$ of Kerr black hole surrounded by a cloud of strings in Rastall gravity are smaller than those of the Kerr black hole surrounded by a cloud of strings in GR for $\beta \in (-2, -\frac{1}{2})$. The reason the curves are divergent near $\beta=-\frac{1}{2}$ in Figs. \ref{rd} and \ref{add} is that the values of obtained photon sphere radii are larger and the properties of expressions $R_s$ and $A$.

In Fig. \ref{dg}, we display the contour plots of $R_{S}$ and $\delta_S$ as well as $A$ and $D$ at inclinations for $\theta=\frac{\pi}{2}$ when $\beta$ is fixed. Through the figure, we can estimate the parameters of Kerr black hole surrounded by a cloud of strings in Rastall gravity. The intersection point of $R_{S}$($A$) and $\delta_S$($D$) determines the value of $a$ and $a_{0}$.

The form of energy emission rate of black holes is \cite{Singh:2017vfr}
\begin{equation}
\frac{d^2E(\omega)}{d\omega dt}=\frac{2\pi^3R_{s}^2}{e^{\frac{\omega}{T_h}}-1}\omega^3,
\end{equation}
where $\omega$ and $T_h$ are the photon frequency and the Hawking temperature at $r_+$, respectively. The Hawking temperature for a Kerr black hole surrounded by a cloud of strings under Rastall gravity is written as
\begin{equation}
T_\mathrm{H}=\frac{(r_+-M)+2a_0r^{1-\lambda_2}\lambda_1\left(2-\lambda_2\right)}{2\pi\left(r_+^2+a^2\right)}.
\end{equation}

The upper row of Fig. \ref{xe} shows the energy emission rate of the black hole for $\beta>0$. The energy emission rate decreases and the peak shifts to the left with the increase of $a_0$. Moreover, the energy emission rate of the Kerr black hole ($a_0$=0) is greater than that of the Kerr black hole surrounded by a cloud of strings in Rastall gravity. The energy emission rate decreases and the peak moves to the left as $\beta$ increases. The bottom row of Fig. \ref{xe} plots the energy emission rate of the black hole for $\beta<0$. With the increase of $a_{0}$, the energy emission decreases and the peak moves to the left. As $\beta$ increases, the energy emission rate first increases and then decreases. The peak of the energy emission rate first shifts to the right and then to the left. Besides, compared with the Kerr black hole surrounded by a cloud of strings in GR ($\beta$=0), the presence of $\beta$ increases the energy emission rate.

\section{Constrains from EHT observations of \text{M87*}} \label{m87}
 \begin{figure*}[ht!]
\centering
 \includegraphics[width=0.432\textwidth]{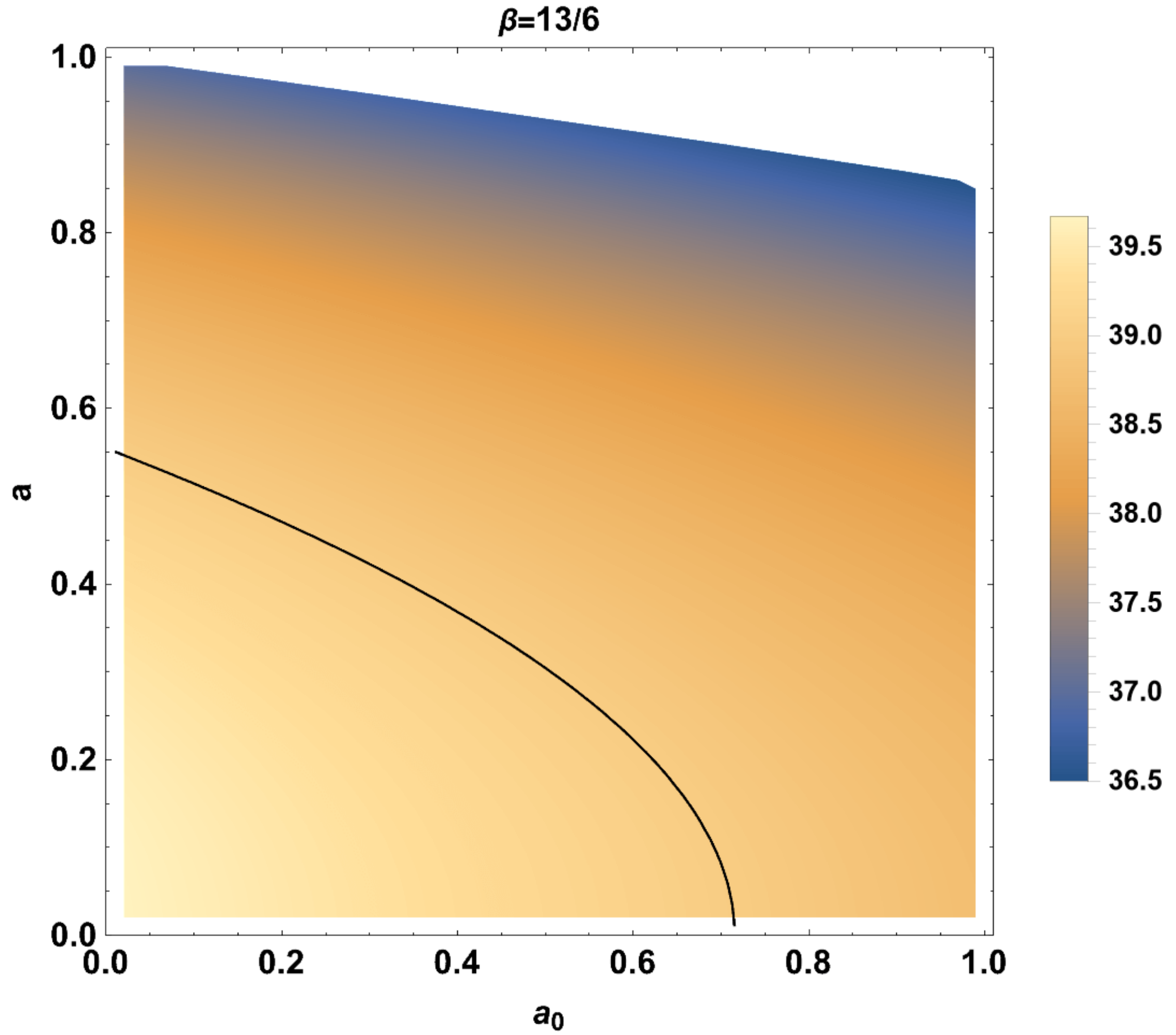}
\includegraphics[width=0.43\textwidth]{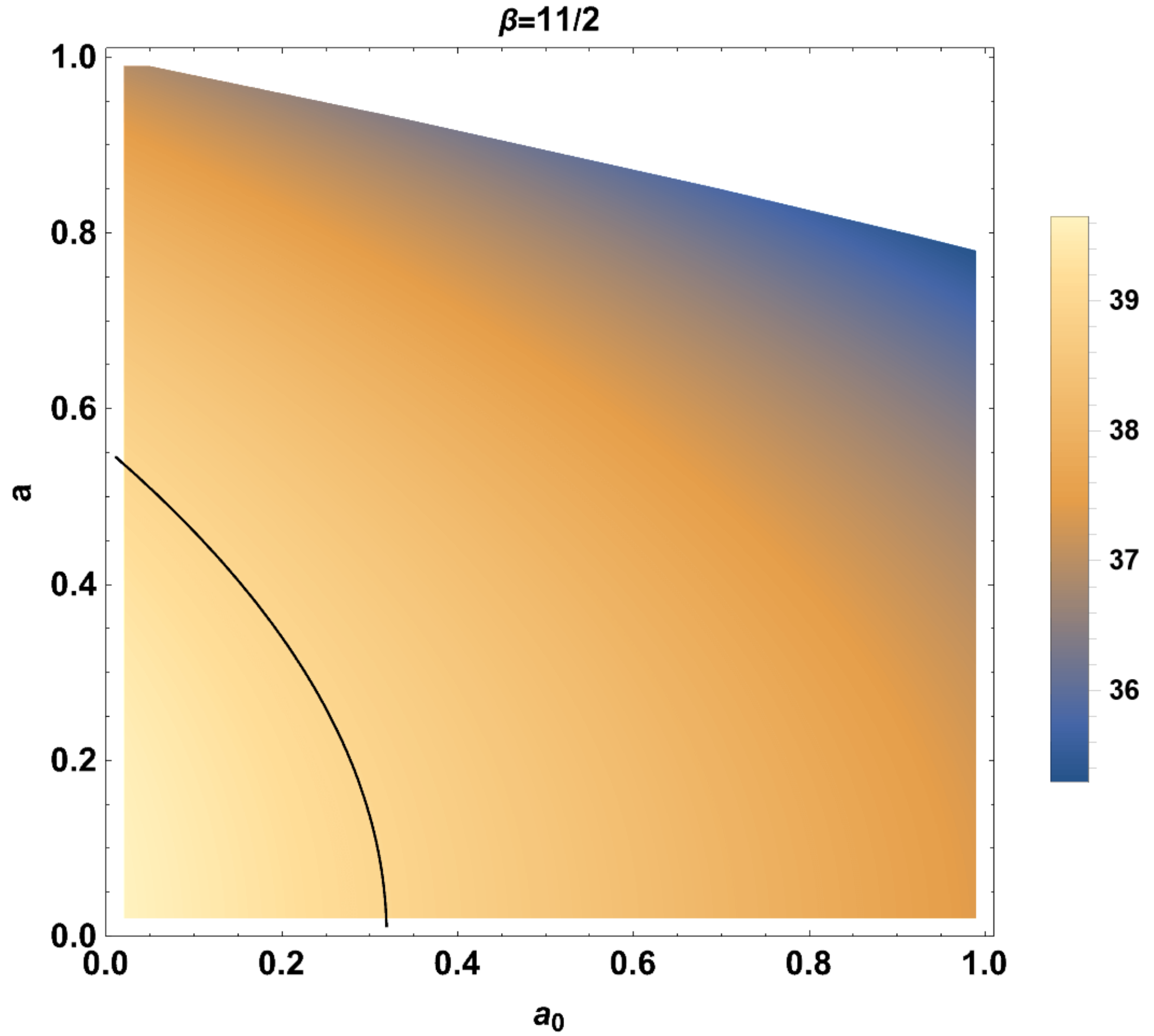}
\includegraphics[width=0.432\textwidth]{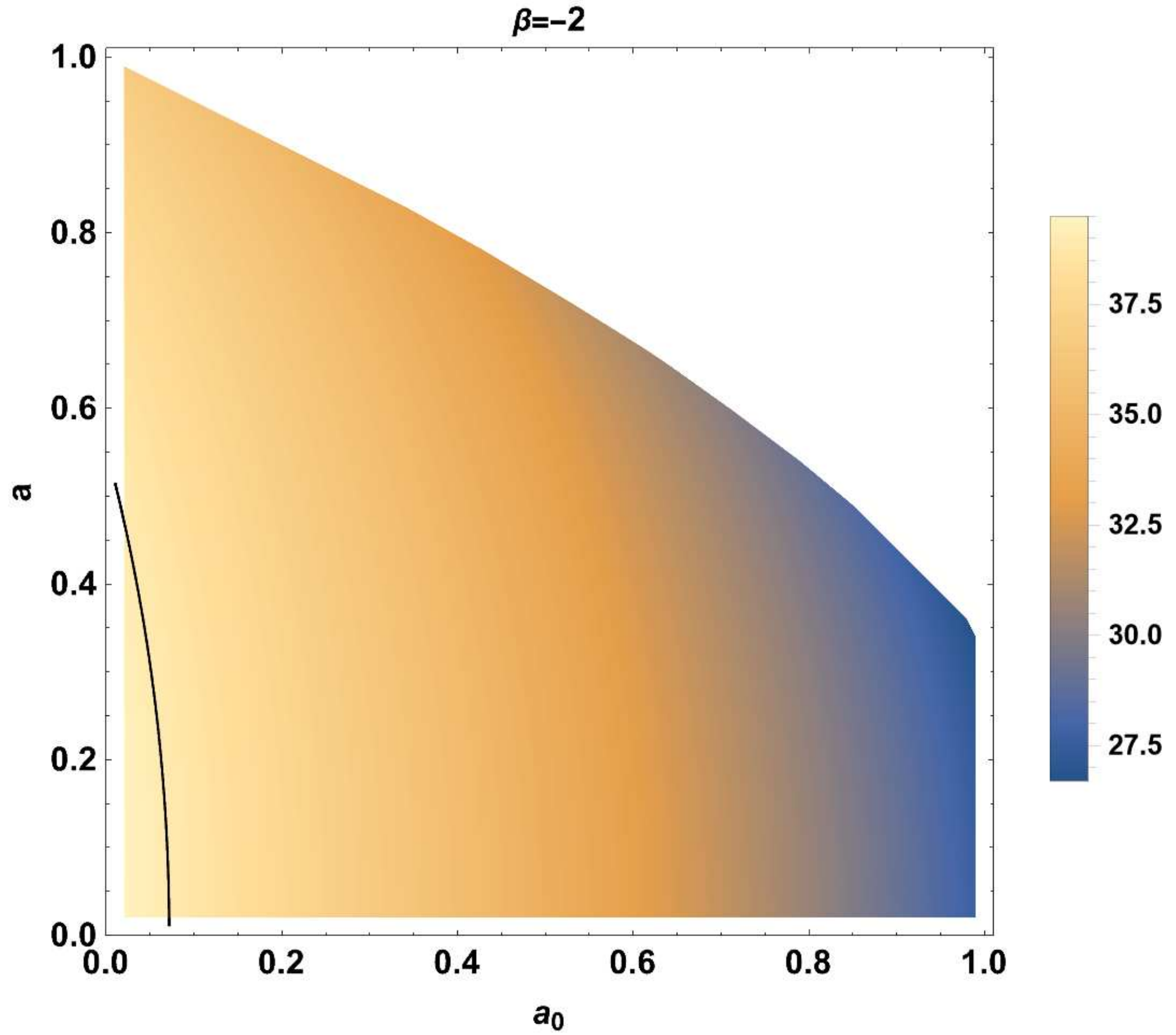}
\includegraphics[width=0.43\textwidth]{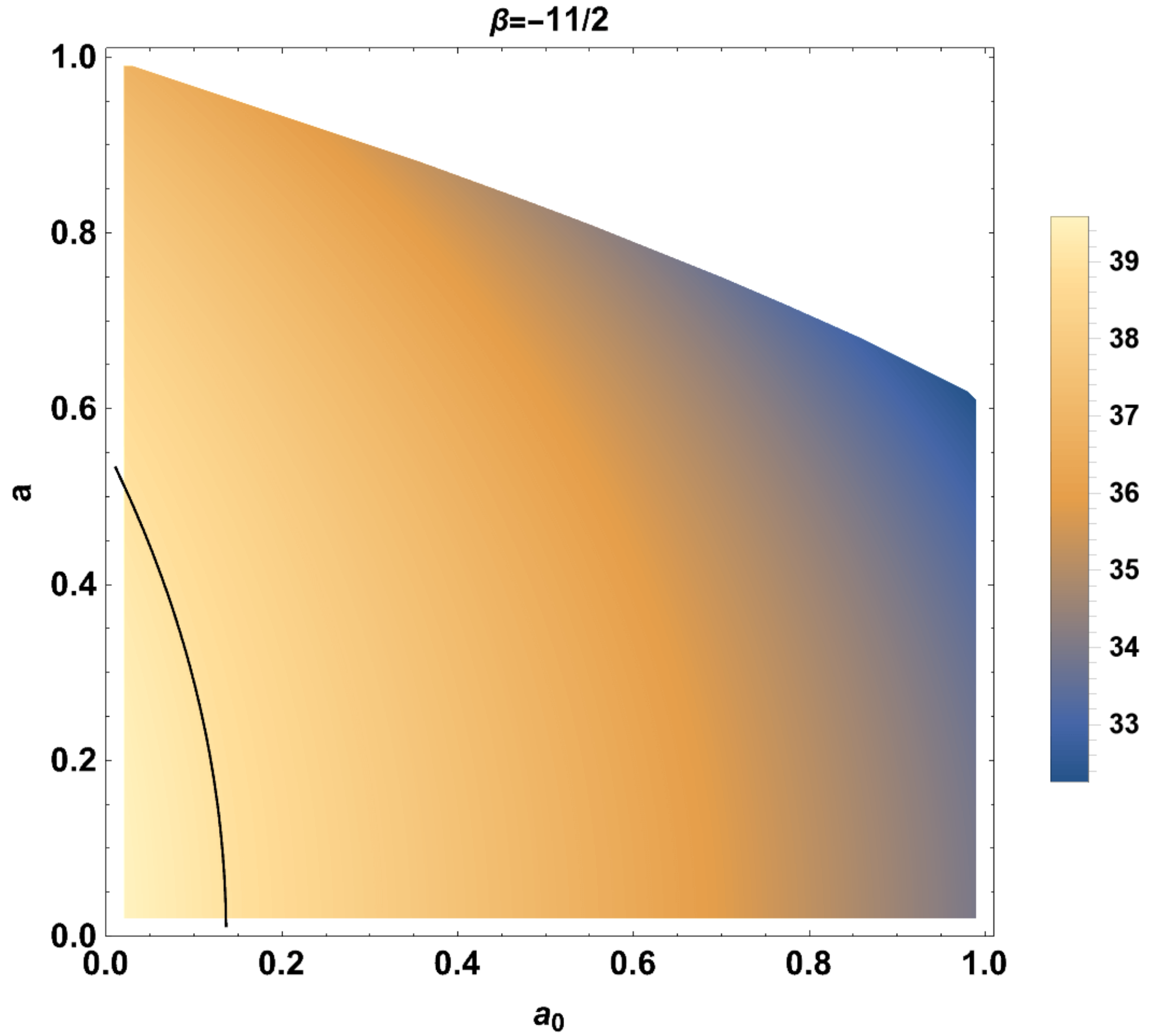}
    \caption{The density plots of angular diameter $\theta_{d}$ as function of $a$ and $a_{0}$ at inclinations for $\theta=17^{\circ}$.}
       \label{thetad}
 \end{figure*}
The EHT collaboration unveiled the supermassive black hole at the center of the giant elliptical galaxy \text{M87} using the EHT, which is a global very long baseline interferometry array. The captured image is a crescent-shaped. And the collaboration extracted some data from the image for preliminary analysis and obtained the circularity deviation $\Delta C\lesssim0.1$, axis ratio $1<D_x\lesssim4/3$, and angular diameter $\theta_{d}=42\pm3 \mu as$. By analyzing these data, they inferred that the captured black hole may be the Kerr black hole in GR. Nevertheless, rotating black holes in MoG may also reconcile with the observational data. Therefore, we view the \text{M87*} as a Kerr black hole surrounded by a cloud of strings in Rastall gravity and use the EHT observations to constrain its parameters. Toward this goal, we calculate the shadow observables $\Delta C$, $D_x$, and $\theta_d$. For the observer's screen, the shadow can be described by polar coordinates ($R(\varphi)$, $\varphi$) with the origin at the center of the shadow $(X_{C},Y_{C})$. Since the black hole spins, the shadow can only move in a direction perpendicular to black holes rotation and is symmetrical along the x-axis. Then, considering the black hole center is at (0,0), the shadow center can be defined as
\begin{equation}
Y_C=0,
\end{equation}
\begin{equation}
 X_C=\frac{(X_r+X_l)}{2}.
\end{equation}

Then the radial coordinate and polar angle of shadow boundary are written as
\begin{equation}
R(\varphi)=\sqrt{(X-X_C)^2+(Y-Y_C)^2},\\
\varphi=\tan^{-1}\left(\frac{Y}{X-X_C}\right).
\end{equation}

The average radius of shadow is given by \cite{Bambi:2019tjh}
\begin{equation}
\bar{R}^{2}=\frac{1}{2\pi}\int_{0}^{2\pi}R^{2}(\varphi)d\varphi.
\end{equation}

With these at hand, we can naturally derive the circularity deviation $\Delta C$, which represents the black hole shadow deviation from a reference circle \cite{Afrin:2021imp},
\begin{equation}
\Delta C=\frac{1}{\bar{R}}\sqrt{\frac{1}{2\pi}\int_{0}^{2\pi}\left(R(\varphi)-\bar{R}\right)^{2}d\varphi}.
\end{equation}

The circular asymmetry in the \text{M87*} shadow can also be measured by the axial ratio, which represents the quotient of the major and minor diameters of the shadow. The expression of the axis ratio is \cite{Banerjee:2019nnj}
\begin{equation}
D_{x}=\frac{1}{D}=\frac{Y_{t}-Y_{b}}{X_{r}-X_{l}}.
\end{equation}

 \begin{figure*}[ht!]
\centering
 \includegraphics[width=0.42\textwidth]{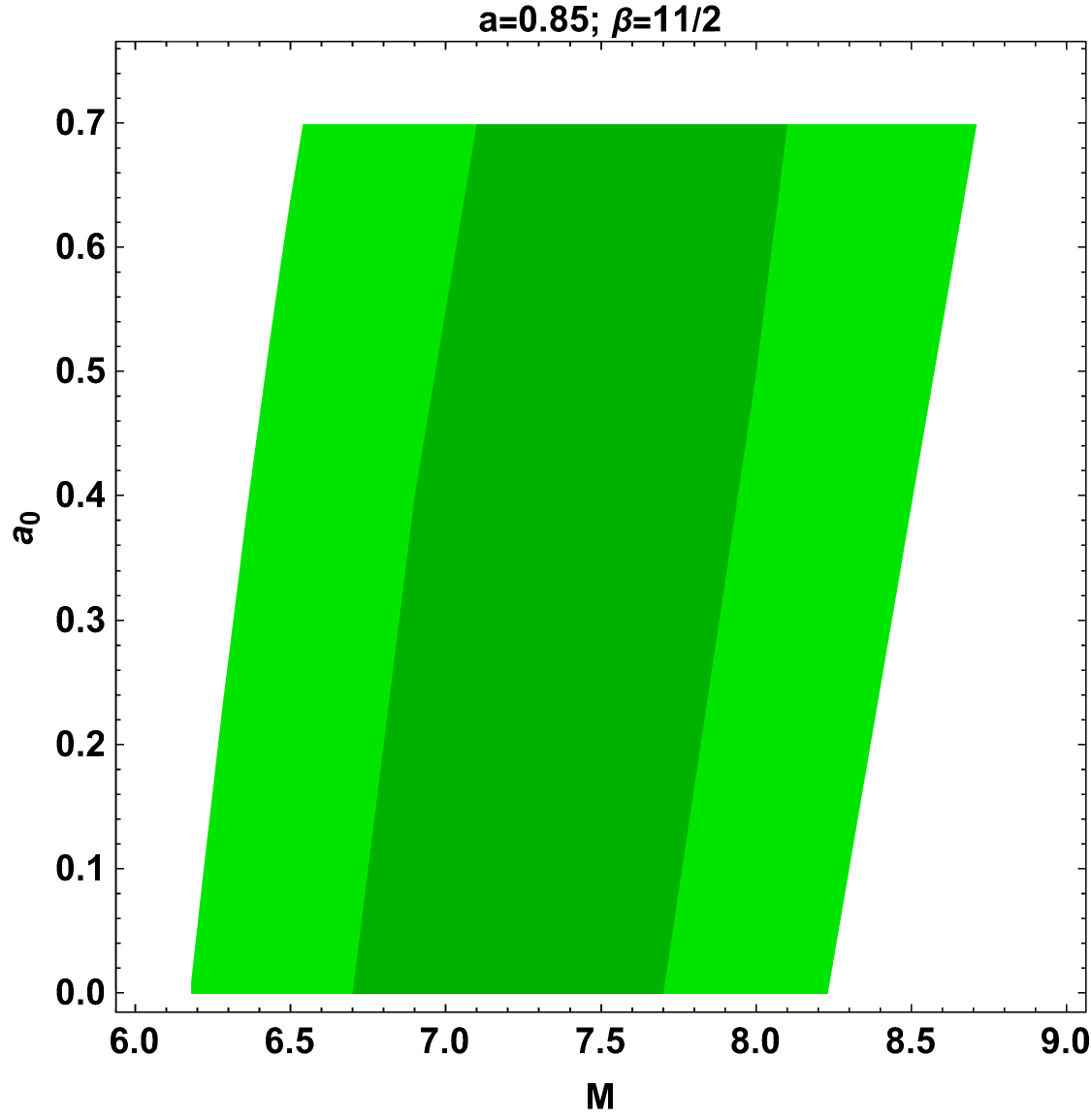}
\includegraphics[width=0.417\textwidth]{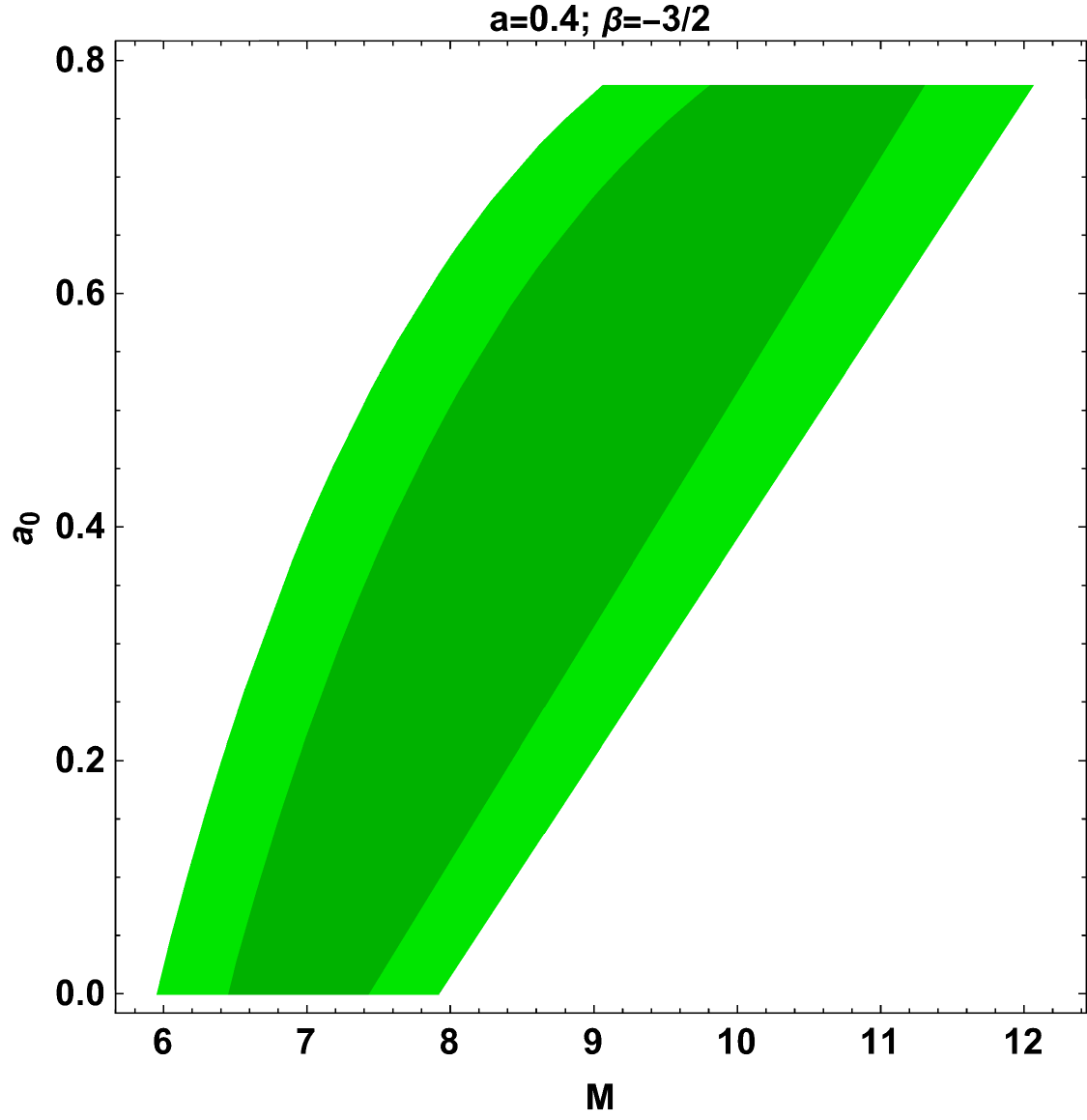}
    \caption{Constraints on parameter $a_0$ and estimated \text{M87*} black hole mass $M(\times10^9M_{\odot})$ using \text{M87*} shadow angular size within 1$\sigma$(dark green region) and 2$\sigma$(light green region).}
       \label{sigma}\end{figure*}

It is the reciprocal of oblateness, which can be seen as another way of defining circular deviation \cite{Meng:2022kjs}. According to EHT observations of \text{M87*}, the axial ratio should be in the range of $1<D_{x}\leq4/3$, which corresponds to $\Delta C\lesssim0.1$.
And the angular diameter of the shadow has the following form \cite{Kumar:2020owy}:
\begin{equation}
\theta_{d}=2\frac{R_{a}}{d},\quad R_{a}=\sqrt{A/\pi},
\end{equation}
in which $A$ can be found from the Eq.~(\ref{AD}); $d$ is the distance between \text{M87*} and the Earth; $R_{a}$ can be called the radius of the shadow area.

Assuming that the supermassive \text{M87*} is a Kerr black hole surrounded by a cloud of strings in Rastall gravity, we set $M=6.5\times10^9M_{\odot}$, $d=16.8Mpc$ and then estimate the observables for the metric. Thereby we can limit some parameters of the black hole using the observables by the EHT such as $\Delta C\lesssim0.1$, $1<D_{x}\leq4/3$, and $\theta_{d}=42\pm3 \mu as$. Furthermore, the paper adopts the inclination angle $\theta=17^{\circ}$, which is estimated in the \text{M87*} image by taking the direction of the relativistic jet into account.

In Fig. \ref{detalc}, we first draw the density plots of the circularity deviation $\Delta C$. Obviously, it shows the shadow of the Kerr black hole surrounded by a cloud of strings in Rastall gravity for various values of $\beta$ ($\beta=\frac{13}{6}$, $\beta=\frac{11}{2}$, $\beta=-2$, $\beta=-\frac{11}{2}$), and all parameter values obey $\Delta C\lesssim0.1$ when $\theta=17^{\circ} $. That is to say, it is difficult to use the EHT observables $\Delta C\lesssim0.1$ to constrain the parameters of the Kerr black hole surrounded by a cloud of strings in Rastall gravity, or to distinguish between GR and Rastall gravity under these circumstances.

Then we display the density plots of axis ratio $D_{x}$ in Fig. \ref{dx}. We find that all parameter spaces $(a_0-a)$ satisfy $1<D_{x}\leq4/3$ for a given $\beta$, which means that it is very consistent with EHT observations. Therefore, we cannot exclude Kerr black holes surrounded by a cloud of strings in Rastall gravity from the observations of the \text{M87*} shadow. Combined with the above analysis, we find that the circularity deviation and axis ratio for EHT bound with respect to M87* allow the entire $(a_0-a)$ space of the Kerr black holes surrounded by a cloud of strings in Rastall gravity under particular parameter $\beta$.

In Fig. \ref{thetad}, we show the density plots of the angular diameter of shadow $\theta_{d}$ for a Kerr black hole surrounded by a cloud of strings under Rastall gravity, and the black line in the figure represents the EHT bound of $39\mu as$. Only the left corner enclosed by the black line which represents the EHT bound for M87* black hole shadow, $\theta_{d}=39\mu as$. It gives the upper limit of the black hole parameters $a$ and $a_{0}$. In the limit, the Kerr black holes surrounded by a cloud of strings in Rastall gravity correspond to the asymmetry of the shadow of M87*. In addition, we can see that $\beta=\frac{11}{2}$ ($\beta=-2$) is more restrictive to $a_{0}$ than $\beta=\frac{13}{6}$ ($\beta=-\frac{11}{2}$).

Futhermore, the restriction on parameter $M$ and $a_0$ for $a=0.85$, $\beta=\frac{11}{2}$ and $a=0.4$, $\beta=-\frac{3}{2}$ by using the 1$\sigma$ and 2$\sigma$ regions of angular diameter in Fig. \ref{sigma}. For $a=0.85$, $\beta=\frac{11}{2}$, 1$\sigma$ bound on mass is $6.69046\times10^9M_{\odot}<M<8.17278\times10^9M_{\odot}$ and 2$\sigma$ bound is $6.17581\times10^9M_{\odot}<M<8.71763\times10^9M_{\odot}$. For $a=0.4$, $\beta=-\frac{3}{2}$, 1$\sigma$ bound on mass is $6.44358\times10^9M_{\odot}<M<11.399\times10^9M_{\odot}$ and 2$\sigma$ bound is $5.94792\times10^9M_{\odot}<M<12.1589\times10^9M_{\odot}$.

\section{Conclusions} \label{Conclusions}

EHT captured the shadow image of \text{M87*} showing a crescent-shaped, which corresponds to the observations of the Kerr black hole predicted in GR. But for now, these observations do not rule out alternatives to Kerr black holes or MoG. Inspired by this, we have considered a Kerr black hole surrounded by a cloud of strings under Rastall gravity in this paper. It introduces additional string parameter $a_0$ and parameter $\beta$ compared with the Kerr black hole. The presence of the parameters causes the black hole to deviate from the Kerr black hole and provides a richer structure. From Fig. \ref{derta}, the increase of $a_0$ and positive $\beta$ reduces the event horizon radius. The change of negative $\beta$ has a more complicated impact on the event horizon. The black holes do not exist at $\beta \in(-0.8891, -\frac{1}{2})$ for $a=0.3$, $a_{0}=0.5$. The influence of $a_0$ and $\beta$ on the ergoregions and photon regions is shown in Fig. \ref{dc} and \ref{xc}. Moreover, we have investigated the influence of $a_0$, $\beta$ on the shadow of the black hole when the observer is at an infinite distance. As $a$ and $a_0$ increase, the deformation of the black hole shadow increases. The black hole shadow size decreases with the increase of positive $\beta$. As negative $\beta$ increases, the size of black hole shadow initially decreases, then increases, followed by another decrease, and finally increases again.

In addition, we have shown the reference circle radius, distortion, shadow area, and oblateness of the shadow in Fig. \ref{rd} and \ref{add}. With the increase of $a_{0}$ and $\beta$, the reference circle radius, shadow area and oblateness decrease, while distortion increases, for $\beta>0$. Owing to the existence of discontinuity point ($\beta=-\frac{1}{2}$), the situation is complex when $\beta<0$. Specifically, the reference circle radius, shadow area and oblateness decrease with the increase in $a_{0}$ when $\beta \in (-2, -\frac{1}{2})$, but increase when $\beta \in (-\frac{1}{2}, 0)$. For the distortion, it increases with the increase of $a_{0}$ in the region of ($-2$, $-\frac{1}{2}$), but decreases in the region of $\beta \in (-\frac{1}{2}, 0)$. With the increase of $\beta$, the reference circle radius, shadow area and oblateness first decrease and then increase in the region of $\beta \in (-\frac{1}{2}, 0)$, while the distortion first increases and then decrease. When $\beta \in (-2, -\frac{1}{2})$, the reference circle radius, shadow area and oblateness decrease with the increase of $\beta$, while distortion increase. And it is also shown that shadow observables can be used to estimate the parameters of Kerr black hole surrounded by a cloud of strings in Rastall gravity.

Moreover, we have shown the energy emission rate of this black hole in Fig. \ref{xe}. With the increase of $a_0$, the energy emission rate decreases when other parameters are fixed. As positive $\beta$ increases, the energy emission rate decreases. In other words, the lifetime of the black hole increases with the increase in $a_0$ and positive $\beta$. Furthermore, the energy emission rate first increases and then decreases as negative $\beta$ increases.

Finally, we have treated \text{M87*} as a Kerr black hole surrounded by a cloud of strings under Rastall gravity, and used EHT data such as circularity deviation, axial ratio, and angular diameter to constrain the black hole parameters in Figs. \ref{detalc}, \ref{dx} and \ref{thetad}. In this paper, we have explored the inclination angle of $17^{\circ}$ which is the inclination angle by taking the orientation of the relativistic jets into account in the \text{M87*} image. We have found that circularity deviation $\Delta C\lesssim0.1$ is satisfied in the entire parameter space. The angular diameter satisfies the observation of EHT within a certain area. And when $\beta$ is larger, the parameter restrictions are stronger. The axis ratio $D_x$ also satisfies $1<D_x\lesssim4/3$ in the whole space. In this way, we have constrained the parameters of a Kerr black hole surrounded by a cloud of strings under Rastall gravity by observations from the EHT. Therefore, \text{M87*} shadow observations do not completely exclude Kerr black hole surrounded by a cloud of strings under Rastall gravity.  That is to say, under some parameter values, a Kerr black hole surrounded by a cloud of strings under Rastall gravity may be a candidate for an astrophysical black hole. The deviation between the angular diameter of a Kerr black hole and the angular diameter of a Kerr black hole surrounded by a cloud of strings under Rastall gravity is tiny. Therefore, judging the current resolution, it is difficult to distinguish between Kerr black holes surrounded by a cloud of strings under Rastall gravity and Kerr black holes. We may anticipate the development of next-generation observational instruments.

\section*{Declaration of competing interest}
The authors declare that they have no known competing financial interests or personal relationships that could have appeared to influence the work reported in this paper.

\section*{Data availability}
No data was used for the research described in the article.

\section*{Acknowledgments}
This work was supported by Yunnan Fundamental Research Projects (Grant No. 202301AS070029), and Yunnan High-level Talent Training Support Plan Young \& Elite Talents Project (Grant No. YNWR-QNBJ-2018-360).

\end{document}